\documentclass[10pt,a4paper]{article}
\usepackage{amsmath,amssymb,amsfonts}
\usepackage{graphicx}
\usepackage{hyperref}
\usepackage[margin=0.75in]{geometry}
\usepackage{float}
\usepackage{bm}
\usepackage{booktabs}
\usepackage{tabularx}

\usepackage{setspace}
\usepackage{titlesec}
\titlespacing*{\section}{0pt}{8pt}{4pt}
\titlespacing*{\subsection}{0pt}{6pt}{3pt}
\titlespacing*{\subsubsection}{0pt}{4pt}{2pt}

\AtBeginDocument{
  \setlength{\abovedisplayskip}{4pt}
  \setlength{\belowdisplayskip}{4pt}
  \setlength{\abovedisplayshortskip}{2pt}
  \setlength{\belowdisplayshortskip}{2pt}
}

\let\oldbibliography\thebibliography
\renewcommand{\thebibliography}[1]{%
  \oldbibliography{#1}%
  \setlength{\itemsep}{0pt}%
  \setlength{\parskip}{0pt}%
}

\title{\texttt{amerta}: A Python Library for Idealized 1D Saint--Venant
Dam-Break Simulation}

\author{
Dasapta E. Irawan$^{1}$,
Sandy H. S. Herho\,$^{2,3,*}$,
Iwan P. Anwar$^{4}$,\\
Faruq Khadami$^{4}$,
Astyka Pamumpuni$^{1}$,
Rendy D. Kartiko$^{1}$,\\
Edi Riawan$^{5}$,
Rusmawan Suwarman$^{5}$,
and Deny J. Puradimaja$^{1}$
}

\date{}

\def\Address{$^{1}$Applied Geology Research Group, Bandung Institute of
Technology, Bandung, Indonesia\\
$^{2}$Department of Earth and Planetary Sciences, University of
California, Riverside, CA, USA\\
$^{3}$Center for Agrarian Studies, Bandung Institute of Technology,
Bandung, Indonesia\\
$^{4}$Applied and Environmental Oceanography Research Group, Bandung
Institute of Technology, Bandung, Indonesia\\
$^{5}$Atmospheric Science Research Group, Bandung Institute of
Technology, Bandung, Indonesia}

\def\corrAuthor{Corresponding Author}
\def\corrEmail{sh001@ucr.edu}

\begin{document}
\maketitle

\begin{center}
\small
\Address\\[2pt]
$^{*}$\corrAuthor: \href{mailto:\corrEmail}{\corrEmail}
\end{center}

\begin{abstract}
\noindent
The Saint--Venant shallow water equations (SWE) govern
depth-integrated free-surface flows arising in dam-break inundation,
flood routing, tsunami runup, and estuarine tidal dynamics.
Closed-form analytical solutions exist only for highly idealized
Riemann configurations, making rigorously verified numerical solvers
essential. This work presents \texttt{amerta}, an open-source Python
library that solves the one-dimensional frictionless Saint--Venant
system on a uniform Cartesian grid using Monotone Upstream-centered
Schemes for Conservation Laws (MUSCL) reconstruction with a minmod
slope limiter, the Harten--Lax--van Leer--Contact (HLLC) approximate
Riemann solver, and two-stage strong-stability-preserving
Runge--Kutta (SSP-RK) time integration. Numba just-in-time (JIT)
compilation accelerates the performance-critical kernels. The solver
is verified end-to-end against the four canonical Riemann
configurations: wet-bed dam break, dry-bed dam break, double
rarefaction, and double shock. A six-component post-processing
pipeline quantifies space-time topology, final-time error norms
with empirical quantile decomposition, self-similarity collapse onto
the analytical Riemann fan, integral-norm evolution,
boundary-flux-corrected mass and energy diagnostics, and phase-plane
analysis against analytical wave curves. The implementation
conserves discrete mass to floating-point precision, satisfies
discrete entropy admissibility identically, and reproduces all four
analytical wave-curve geometries to within sub-centimetre accuracy
in the depth-velocity phase plane. The complete source code,
analytical-solution evaluators, post-processing scripts, and
Network Common Data Format (NetCDF) archives are released under the
MIT license.
\end{abstract}


\section{Introduction}

The shallow water equations (SWE) describe the depth-integrated
dynamics of incompressible free-surface flows under the
hydrostatic approximation~\cite{ref-svc1871}. The system arises
naturally across a broad range of geophysical and hydraulic
applications in which the characteristic horizontal length scale of
the flow greatly exceeds the vertical scale, including dam-break
and flood-routing analyses in rivers and artificial
channels~\cite{ref-toro2001,ref-leveque2002}, tsunami propagation
and runup over continental shelves, storm-surge inundation in
coastal estuaries, and tidal dynamics in shallow basins. The
reduction from the three-dimensional incompressible Navier--Stokes
equations to the depth-averaged form eliminates the vertical
structure of the velocity field at the cost of restricting
attention to flows whose vertical accelerations remain small
compared to gravity, a balance routinely satisfied in
geophysically and hydraulically relevant
regimes~\cite{ref-vreugdenhil1994,ref-batchelor1967}.

The one-dimensional Saint--Venant system in conservation form
constitutes a strictly hyperbolic system of conservation laws
whose structure is closely analogous to the compressible Euler
equations of gas dynamics~\cite{ref-leveque2002,ref-toro2001}. The
eigenstructure of the flux Jacobian admits two genuinely nonlinear
characteristic fields whose propagation speeds depend on the local
flow velocity and the long-wave celerity, producing wave
structures composed of combinations of rarefactions and shocks
separated by a constant intermediate state. The piecewise-constant
Riemann problem for the shallow water system admits closed-form
analytical solutions in the dry-bed
configuration~\cite{ref-ritter1892} and in the wet-bed
configuration~\cite{ref-stoker1957}; together with the symmetric
double-rarefaction and double-shock variants, these four canonical
problems exhaust the elementary wave structures admissible to the
source-free system in the absence of bed slope and friction, and
constitute the mathematical benchmarks against which any
consistent shallow water solver must be
verified~\cite{ref-lax1957,ref-kruzhkov1970}.

The numerical solution of the Saint--Venant system in regimes
where shocks, dry beds, and steep gradients arise simultaneously
has driven the development of a substantial literature on
high-resolution shock-capturing schemes. Godunov-type
finite-volume methods, in which the inter-cell flux is
reconstructed from the local solution of an exact or approximate
Riemann problem, provide the standard
framework~\cite{ref-leveque2002}, with the Harten--Lax--van Leer
(HLL) family of approximate Riemann solvers and the contact-restoring
Harten--Lax--van Leer--Contact (HLLC) variant delivering both
robustness and computational
economy~\cite{ref-hartenlaxleer1983,ref-einfeldt1988,ref-toro1994}.
Second-order spatial accuracy in smooth regions is recovered
through Monotone Upstream-centered Schemes for Conservation Laws
(MUSCL)~\cite{ref-vanleer1979} combined with
total-variation-diminishing (TVD) slope
limiters~\cite{ref-harten1983,ref-sweby1984}, and explicit time
integration through strong-stability-preserving Runge--Kutta
(SSP-RK) methods preserves the discrete entropy and TVD properties
of the spatial
discretization~\cite{ref-shuosher1988,ref-gottlieb2001}. Further
extensions to wet-and-drying interfaces, variable bathymetry, and
steady-state preservation have produced well-balanced and
augmented Riemann-solver formulations capable of handling the
transcritical and dry-bed regimes characteristic of realistic
shallow flows~\cite{ref-audusse2004,ref-george2008}, and a
standardized benchmark library of analytical solutions enables
systematic cross-validation across solver
implementations~\cite{ref-delestre2013}.

The wider availability of just-in-time (JIT) compiled scientific
Python libraries has substantially altered the implementation
landscape for computational fluid dynamics (CFD) by enabling
research-grade performance from interpreted high-level languages
while preserving readability and
portability~\cite{ref-numpy,ref-scipy,ref-numba,ref-matplotlib}.
Recent open-source contributions from the present research group
illustrate this trend across distinct branches of geophysical
fluid mechanics: an idealized two-dimensional incompressible
Kelvin--Helmholtz (KH) instability solver for stratified shear
flows that combines a fractional-step projection method with a
spectral Poisson solution via the fast sine transform
(FST)~\cite{ref-kh2d2025}; a one-dimensional linearized
Saint--Venant framework for wave attenuation through coastal
vegetation that demonstrates the same acceleration strategy in a
fundamentally different physical regime of incident-wave
dissipation through plant drag~\cite{ref-waveatt2026}; and a
Fourier pseudo-spectral solver for the Korteweg--de Vries (KdV)
equation with adaptive eighth-order Runge--Kutta time integration
that quantifies multi-soliton dynamics through complementary
information-theoretic and recurrence-quantification
diagnostics~\cite{ref-sangkuriang2026}. Across these
implementations the design philosophy is consistent: a focused
physical model, a numerical scheme appropriate to the regime,
JIT-compiled performance-critical kernels, and self-describing
Network Common Data Format (NetCDF) outputs paired with
reproducible analysis scripts.

Despite this substantial methodological literature, accessible
reference implementations of canonical Saint--Venant Riemann
solvers that combine end-to-end verification against closed-form
analytical solutions with a transparent, well-documented Python
codebase remain limited. The present work addresses this gap with
\texttt{amerta}, an open-source Python library that implements the
MUSCL--HLLC finite-volume scheme with two-stage SSP-RK time
integration for the one-dimensional frictionless Saint--Venant
system on a uniform Cartesian grid, with Numba JIT compilation of
the performance-critical kernels and self-describing NetCDF
output. The numerical core is deliberately restricted to the
frictionless prismatic-channel limit so that solver behavior can be
evaluated end-to-end against the closed-form Ritter, Stoker,
double-rarefaction, and double-shock Riemann solutions; this
idealization is not a claim of physical generality but the precise
condition under which mathematical verification against exact
solutions is feasible. The four canonical Riemann configurations
are provided as default verification cases, the corresponding
analytical-solution evaluators are bundled in the same package,
and a six-component post-processing pipeline quantifies space-time
field topology, final-time validation against analytical Riemann
solutions, self-similarity collapse onto the analytical Riemann
fan, time evolution of integral error norms in depth, discharge,
and wet-cell velocity, boundary-flux-corrected mass and energy
diagnostics, and phase-plane analysis against analytical wave
curves. The library is released under the MIT license with the
complete source code, configuration files, post-processing
scripts, and NetCDF archives of all benchmark runs publicly
available in the project repositories.
\section{Methods}

\subsection{Model Description}

The one-dimensional Saint--Venant system governs the
depth-averaged dynamics of a free-surface flow over a rigid, impermeable,
horizontal bed, and constitutes the standard shallow-water model used in
dam-break and flood-routing analysis~\cite{ref-toro2001,ref-leveque2002}.
We derive this system from the three-dimensional incompressible
Navier--Stokes equations through three explicit physical assumptions,
fixing the regime of validity of the conservative form that the
\texttt{amerta} solver discretizes~\cite{ref-stoker1957,
ref-vreugdenhil1994}.

Consider an incompressible Newtonian fluid of constant density
$\rho \in (0, \infty)$ [kg\,m$^{-3}$] and constant kinematic viscosity
$\nu \in [0, \infty)$ [m$^{2}$\,s$^{-1}$] occupying the time-dependent
domain
\begin{equation}
\Omega(t) \;=\; \bigl\{\,\mathbf{x} \in \mathbb{R}^{3} \,:\,
0 < x_{3} < h(x_{1}, x_{2}, t)\,\bigr\},
\qquad t \in [0, T],
\label{eq:domain}
\end{equation}
where $h \in C^{1}\!\bigl(\mathbb{R}^{2} \times [0, T]\bigr)$, with
$h > 0$, denotes the local water depth measured from the rigid bed at
$x_{3} = 0$ to the free surface $x_{3} = h$. We adopt Cartesian
coordinates $(x_{1}, x_{2}, x_{3}) \equiv (x, y, z)$ with orthonormal
basis $\{\mathbf{e}_{i}\}_{i=1}^{3}$, and write the velocity field
as $\mathbf{u} = u_{i}\,\mathbf{e}_{i}$ with components
$(u_{1}, u_{2}, u_{3}) \equiv (u, v, w)$ [m\,s$^{-1}$], assumed to lie
in $C^{1}\!\bigl(\Omega(t) \times [0, T]\bigr)$. The Einstein summation
convention applies to repeated lower indices throughout. Let $p:
\Omega(t) \times [0,T] \to \mathbb{R}$ denote the pressure field [Pa],
and let $g \in (0, \infty)$ [m\,s$^{-2}$] denote the magnitude of the
gravitational acceleration, which acts in the $-\mathbf{e}_{3}$
direction.

Conservation of mass and balance of linear momentum in $\Omega(t)$ are
governed by the incompressible Navier--Stokes
equations~\cite{ref-batchelor1967,ref-landau1987},
\begin{equation}
\frac{\partial u_{i}}{\partial x_{i}} \;=\; 0,
\label{eq:ns_continuity}
\end{equation}
\begin{equation}
\rho\!\left(
\frac{\partial u_{i}}{\partial t}
+ u_{j}\,\frac{\partial u_{i}}{\partial x_{j}}
\right)
\;=\;
-\,\frac{\partial p}{\partial x_{i}}
+ \frac{\partial \tau_{ij}}{\partial x_{j}}
+ \rho\,g_{i},
\qquad i \in \{1,2,3\},
\label{eq:ns_momentum}
\end{equation}
where $g_{i} = -g\,\delta_{i3}$ with $\delta_{ij}$ the Kronecker delta,
and the viscous stress tensor $\tau_{ij}$ [Pa] satisfies the Newtonian
constitutive relation
\begin{equation}
\tau_{ij} \;=\;
\rho\,\nu\!\left(
\frac{\partial u_{i}}{\partial x_{j}}
+ \frac{\partial u_{j}}{\partial x_{i}}
\right),
\label{eq:newtonian_stress}
\end{equation}
whose trace
$\tau_{ii} = 2\rho\,\nu\,\partial u_{i}/\partial x_{i}$
vanishes identically by~\eqref{eq:ns_continuity}, so that $\tau_{ij}$
is automatically deviatoric.

The system~\eqref{eq:ns_continuity}, \eqref{eq:ns_momentum} is closed
by kinematic and dynamic conditions on $\partial \Omega(t)$~\cite{
ref-stoker1957,ref-vreugdenhil1994}. Impermeability of the rigid bed at
$z = 0$ requires
\begin{equation}
w\,\big|_{z=0} \;=\; 0.
\label{eq:bc_bed}
\end{equation}
The kinematic free-surface condition, which expresses the material
invariance of the surface $z = h(x, y, t)$ under the flow, reads
\begin{equation}
\frac{\partial h}{\partial t}
+ u\,\big|_{z=h}\,\frac{\partial h}{\partial x}
+ v\,\big|_{z=h}\,\frac{\partial h}{\partial y}
\;=\; w\,\big|_{z=h}.
\label{eq:bc_kinematic}
\end{equation}
The dynamic condition, neglecting surface tension and adopting the
pressure gauge $p \mapsto p - p_{\mathrm{atm}}$ with
$p_{\mathrm{atm}} \in \mathbb{R}$ [Pa] the constant atmospheric
pressure, reduces to
\begin{equation}
p\,\big|_{z=h} \;=\; 0.
\label{eq:bc_dynamic}
\end{equation}

The reduction to the Saint--Venant system proceeds under three
assumptions, which we state and apply in sequence.

\textit{Assumption~1 (plane flow).} We restrict attention to plane
motion in the $(x, z)$ half-plane along a wide prismatic channel by
setting $v \equiv 0$ and $\partial(\cdot)/\partial y \equiv 0$.
This restriction is appropriate to dam-break problems in long, straight
reaches in which transverse variation is
negligible~\cite{ref-stoker1957}. The
system~\eqref{eq:ns_continuity}, \eqref{eq:ns_momentum} then collapses
to
\begin{equation}
\frac{\partial u}{\partial x} + \frac{\partial w}{\partial z} \;=\; 0,
\label{eq:ns_2d_continuity}
\end{equation}
\begin{equation}
\rho\!\left(
\frac{\partial u}{\partial t}
+ u\,\frac{\partial u}{\partial x}
+ w\,\frac{\partial u}{\partial z}
\right)
\;=\;
-\,\frac{\partial p}{\partial x}
+ \frac{\partial \tau_{1j}}{\partial x_{j}},
\label{eq:ns_2d_momx}
\end{equation}
\begin{equation}
\rho\!\left(
\frac{\partial w}{\partial t}
+ u\,\frac{\partial w}{\partial x}
+ w\,\frac{\partial w}{\partial z}
\right)
\;=\;
-\,\frac{\partial p}{\partial z}
+ \frac{\partial \tau_{3j}}{\partial x_{j}}
- \rho\,g.
\label{eq:ns_2d_momz}
\end{equation}

\textit{Assumption~2 (shallow-water scaling and inviscid limit).}
Introduce a horizontal length scale $\mathcal{L}$ [m], a vertical length
scale $\mathcal{H}$ [m], and a horizontal velocity scale
$\mathcal{U} = \sqrt{g\,\mathcal{H}}$ [m\,s$^{-1}$]. Define the
shallowness parameter and the Reynolds number by
\begin{equation}
\beta \;\equiv\; \!\left(\frac{\mathcal{H}}{\mathcal{L}}\right)^{\!2},
\qquad \beta \ll 1,
\qquad
\mathrm{Re} \;\equiv\; \frac{\mathcal{U}\,\mathcal{L}}{\nu},
\label{eq:shallowness}
\end{equation}
respectively. Non-dimensionalizing
\eqref{eq:ns_2d_continuity}--\eqref{eq:ns_2d_momz} with these scales
shows that all inertial contributions to the vertical momentum
equation~\eqref{eq:ns_2d_momz} are of relative order
$\mathcal{O}(\beta)$ against gravity~\cite{ref-vreugdenhil1994}. We
further adopt the frictionless inviscid limit
$\nu \to 0$, equivalently $\mathrm{Re} \to \infty$ and
$\tau_{ij} \to 0$, consistent with the configuration of the
\texttt{amerta} solver, in which bottom friction and turbulent closure
are excluded so that the canonical Riemann
problems~\cite{ref-ritter1892,ref-stoker1957} admit exact analytical
solutions for solver verification. At leading order in $\beta$, the
vertical balance~\eqref{eq:ns_2d_momz} reduces to the hydrostatic
relation
\begin{equation}
\frac{\partial p}{\partial z} \;=\; -\rho\,g.
\label{eq:hydrostatic_balance}
\end{equation}
Integrating~\eqref{eq:hydrostatic_balance} from a generic elevation $z$
to the free surface $z = h(x, t)$ and applying~\eqref{eq:bc_dynamic}
yields
\begin{equation}
p(x, z, t) \;=\; \rho\,g\,\bigl[h(x, t) - z\bigr],
\label{eq:hydrostatic_pressure}
\end{equation}
from which the horizontal pressure gradient
\begin{equation}
\frac{\partial p}{\partial x} \;=\;
\rho\,g\,\frac{\partial h}{\partial x}
\label{eq:pressure_gradient}
\end{equation}
is independent of $z$.

\textit{Assumption~3 (depth-uniform horizontal velocity).} Since the
horizontal pressure gradient~\eqref{eq:pressure_gradient} carries no
explicit $z$-dependence and viscous coupling between fluid layers has
been suppressed by Assumption~2, equation~\eqref{eq:ns_2d_momx} admits
solutions in which the horizontal velocity is independent of $z$. We
define the depth-averaged horizontal velocity
\begin{equation}
\bar{u}(x, t) \;\equiv\;
\frac{1}{h(x, t)}\,\int_{0}^{h(x, t)} u(x, z, t)\,\mathrm{d}z,
\label{eq:depth_average}
\end{equation}
and, consistent with the leading-order asymptotics, identify
$u(x, z, t) = \bar{u}(x, t)$ for all $z \in [0,\,h(x, t)]$; the overbar
is dropped hereafter. This identification is exact under the
hydrostatic, inviscid scaling and defines the Saint--Venant closure of
the depth-averaged dynamics~\cite{ref-stoker1957,ref-vreugdenhil1994}.

The depth-averaged conservation laws follow from vertical integration
of~\eqref{eq:ns_2d_continuity} and~\eqref{eq:ns_2d_momx} together with
the Leibniz rule for differentiation under the integral sign with a
moving upper limit~\cite{ref-flanders1973}: for any
$f \in C^{1}\!\bigl(\mathbb{R} \times [0,\,h(x, t)] \times [0, T]\bigr)$,
\begin{equation}
\frac{\partial}{\partial x}\!\int_{0}^{h(x,t)}\! f(x, z, t)\,\mathrm{d}z
\;=\;
\int_{0}^{h(x,t)}\!
\frac{\partial f}{\partial x}(x, z, t)\,\mathrm{d}z
\;+\;
f\!\bigl(x, h(x, t), t\bigr)\,\frac{\partial h}{\partial x}(x, t).
\label{eq:leibniz}
\end{equation}

Integrating~\eqref{eq:ns_2d_continuity} in $z$ from $0$ to
$h(x, t)$ produces
\begin{equation}
\int_{0}^{h}\!\frac{\partial u}{\partial x}\,\mathrm{d}z
\;+\;
\int_{0}^{h}\!\frac{\partial w}{\partial z}\,\mathrm{d}z
\;=\; 0.
\label{eq:continuity_int_raw}
\end{equation}
The second integral evaluates by the fundamental theorem of calculus to
$w\,\big|_{z=h} - w\,\big|_{z=0} = w\,\big|_{z=h}$, where the second
equality uses~\eqref{eq:bc_bed}. Applying~\eqref{eq:leibniz} to the
first integral with $f = u$ and invoking
$\int_{0}^{h}\! u\,\mathrm{d}z = h u$ from the depth-uniformity
established in Assumption~3 gives
\begin{equation}
\int_{0}^{h}\!\frac{\partial u}{\partial x}\,\mathrm{d}z
\;=\;
\frac{\partial (h u)}{\partial x}
\;-\;
u\,\big|_{z=h}\,\frac{\partial h}{\partial x}
\;=\;
\frac{\partial (h u)}{\partial x}
\;-\;
u\,\frac{\partial h}{\partial x}.
\label{eq:continuity_leibniz}
\end{equation}
Substituting~\eqref{eq:continuity_leibniz} and the kinematic
condition~\eqref{eq:bc_kinematic} (with $v \equiv 0$, $\partial/\partial
y \equiv 0$) into~\eqref{eq:continuity_int_raw} yields
\begin{equation}
\frac{\partial (h u)}{\partial x}
\;-\;
u\,\frac{\partial h}{\partial x}
\;+\;
\frac{\partial h}{\partial t}
\;+\;
u\,\frac{\partial h}{\partial x}
\;=\; 0,
\label{eq:mass_intermediate}
\end{equation}
in which the $u\,\partial_{x} h$ terms cancel exactly. The depth-averaged
mass-conservation law therefore reads
\begin{equation}
\frac{\partial h}{\partial t}
\;+\;
\frac{\partial (h u)}{\partial x}
\;=\; 0,
\label{eq:sv_mass}
\end{equation}
expressing conservation of water volume per unit channel width.

Integrating~\eqref{eq:ns_2d_momx} in $z$ from $0$ to $h(x, t)$ with
$\tau_{ij} = 0$ and dividing through by $\rho$,
\begin{equation}
\int_{0}^{h}\!\frac{\partial u}{\partial t}\,\mathrm{d}z
+ \int_{0}^{h}\! u\,\frac{\partial u}{\partial x}\,\mathrm{d}z
+ \int_{0}^{h}\! w\,\frac{\partial u}{\partial z}\,\mathrm{d}z
\;=\;
-\,\frac{1}{\rho}\int_{0}^{h}\!\frac{\partial p}{\partial x}\,\mathrm{d}z.
\label{eq:mom_integrated}
\end{equation}
The depth-uniformity of $u$ implies $\partial u/\partial z = 0$
identically, so the third integral on the left-hand side
of~\eqref{eq:mom_integrated} vanishes. The remaining left-hand integrals
evaluate to $h\,\partial u/\partial t$ and $h u\,\partial u/\partial x$,
respectively. The right-hand side reduces to $-g h\,\partial h/\partial
x$ by~\eqref{eq:pressure_gradient}, which is independent of $z$.
Equation~\eqref{eq:mom_integrated} therefore becomes
\begin{equation}
h\,\frac{\partial u}{\partial t}
+ h u\,\frac{\partial u}{\partial x}
+ g h\,\frac{\partial h}{\partial x}
\;=\; 0,
\label{eq:sv_momentum_primitive}
\end{equation}
which is the primitive (non-conservation) form of the Saint--Venant
momentum equation.

To recast~\eqref{eq:sv_momentum_primitive} in conservation form,
multiply the mass-conservation law~\eqref{eq:sv_mass} by $u$ and add the
result to~\eqref{eq:sv_momentum_primitive}. Using the differential
identities
\begin{equation}
u\,\frac{\partial h}{\partial t} + h\,\frac{\partial u}{\partial t}
\;=\; \frac{\partial (h u)}{\partial t},
\qquad
u\,\frac{\partial (h u)}{\partial x} + h u\,\frac{\partial u}{\partial x}
\;=\; \frac{\partial (h u^{2})}{\partial x},
\qquad
g h\,\frac{\partial h}{\partial x}
\;=\; \frac{\partial}{\partial x}\!\left(\tfrac{1}{2}\,g h^{2}\right),
\label{eq:product_rules}
\end{equation}
where the first two follow from the Leibniz product rule
and~\eqref{eq:sv_mass}, the result is the conservative momentum balance
\begin{equation}
\frac{\partial (h u)}{\partial t}
\;+\;
\frac{\partial}{\partial x}\!\left(
h u^{2} + \tfrac{1}{2}\,g h^{2}
\right) \;=\; 0,
\label{eq:sv_momentum}
\end{equation}
in which the flux contribution $\tfrac{1}{2}\,g h^{2}$ is the depth
integral of the hydrostatic pressure~\eqref{eq:hydrostatic_pressure}.

Define the open positive-depth state space
$\mathbb{R}^{2}_{+} = \{(h, q) \in \mathbb{R}^{2} : h > 0\}$,
the conserved-variable vector
$\mathbf{U} : \mathbb{R} \times [0, T] \to \mathbb{R}^{2}_{+}$, and
the flux vector $\mathbf{F} : \mathbb{R}^{2}_{+} \to \mathbb{R}^{2}$ by
\begin{equation}
\mathbf{U}(x, t)
\;=\;
\begin{pmatrix} h \\[2pt] q \end{pmatrix},
\qquad
\mathbf{F}(\mathbf{U})
\;=\;
\begin{pmatrix}
q \\[4pt]
\dfrac{q^{2}}{h} + \dfrac{1}{2}\,g\,h^{2}
\end{pmatrix},
\label{eq:sv_state_flux}
\end{equation}
where $q(x, t) \equiv h(x, t)\,u(x, t)$ [m$^{2}$\,s$^{-1}$] is the
specific discharge, i.e., the volumetric flux per unit channel width.
The pair~\eqref{eq:sv_mass},~\eqref{eq:sv_momentum} then admits the
compact hyperbolic conservation-law form
\begin{equation}
\frac{\partial \mathbf{U}}{\partial t}
\;+\;
\frac{\partial \mathbf{F}(\mathbf{U})}{\partial x}
\;=\; \mathbf{0},
\qquad x \in \mathbb{R},\quad t > 0,\quad \mathbf{U}(x, t) \in \mathbb{R}^{2}_{+},
\label{eq:sv_conservative}
\end{equation}
supplemented by initial data $\mathbf{U}(x, 0) = \mathbf{U}_{0}(x)$ with
$\mathbf{U}_{0} : \mathbb{R} \to \mathbb{R}^{2}_{+}$.

The flux Jacobian
$\mathbf{A}(\mathbf{U}) \equiv \partial \mathbf{F}/\partial \mathbf{U}
\in \mathbb{R}^{2 \times 2}$ is obtained by direct differentiation
of~\eqref{eq:sv_state_flux}. Using $\partial F_{1}/\partial h = 0$,
$\partial F_{1}/\partial q = 1$, $\partial F_{2}/\partial h = -q^{2}/h^{2}
+ g h = c^{2} - u^{2}$, and $\partial F_{2}/\partial q = 2 q/h = 2 u$,
we obtain
\begin{equation}
\mathbf{A}(\mathbf{U})
\;=\;
\begin{pmatrix}
0 & 1 \\[3pt]
c^{2} - u^{2} & 2\,u
\end{pmatrix},
\qquad
c(\mathbf{U}) \;\equiv\; \sqrt{g\,h},
\label{eq:sv_jacobian}
\end{equation}
where $c$ [m\,s$^{-1}$] is the local gravity-wave celerity and
$u = q/h$ has been reintroduced for transparency. The characteristic
polynomial
$\det\!\bigl(\mathbf{A}(\mathbf{U}) - \lambda\,\mathbf{I}_{2}\bigr) = 0$
of~\eqref{eq:sv_jacobian} expands as
\begin{equation}
\lambda^{2} \;-\; 2\,u\,\lambda \;+\; (u^{2} - c^{2}) \;=\; 0,
\label{eq:char_poly}
\end{equation}
which factors as $(\lambda - u)^{2} = c^{2}$ and admits the two real
eigenvalues
\begin{equation}
\lambda_{\pm}(\mathbf{U}) \;=\; u \pm c,
\label{eq:sv_eigenvalues}
\end{equation}
distinct whenever $h > 0$, with associated right eigenvectors
\begin{equation}
\mathbf{r}_{\pm}(\mathbf{U}) \;=\;
\begin{pmatrix} 1 \\ u \pm c \end{pmatrix},
\label{eq:sv_eigenvectors}
\end{equation}
obtained by solving
$\bigl(\mathbf{A} - \lambda_{\pm}\,\mathbf{I}_{2}\bigr)\,\mathbf{r}_{\pm}
= \mathbf{0}$ with the normalization $(r_{\pm})_{1} = 1$. The
system~\eqref{eq:sv_conservative} is therefore strictly hyperbolic on
$\mathbb{R}^{2}_{+}$ and degenerates at the dry state $h \to 0$, where
$\lambda_{+}$ and $\lambda_{-}$ coalesce to the single eigenvalue
$u$~\cite{ref-toro2001}. The Riemann invariants $u \pm 2 c$, which are
constant along the characteristic curves
$\mathrm{d}x/\mathrm{d}t = \lambda_{\pm}$, form the analytical basis for
the closed-form solutions of the canonical dam-break Riemann
problems~\cite{ref-ritter1892,ref-stoker1957} that serve as verification
benchmarks for the \texttt{amerta} solver.
Equation~\eqref{eq:sv_conservative}, together with the state space
$\mathbb{R}^{2}_{+}$ and the flux~\eqref{eq:sv_state_flux}, is the
continuum mathematical model that the solver discretizes.

\subsection{Numerical Implementation and Verification Cases}

The hyperbolic conservation law~\eqref{eq:sv_conservative} is
discretized in space by a Godunov-type finite-volume method with
second-order MUSCL reconstruction~\cite{ref-vanleer1979} and an HLLC
approximate Riemann solver~\cite{ref-toro1994}, and in time by the
two-stage SSP-RK scheme of Shu and
Osher~\cite{ref-shuosher1988,ref-gottlieb2001}. The implementation is
deliberately restricted to the frictionless prismatic-channel limit
introduced above, so that the canonical Riemann problems of
Ritter~\cite{ref-ritter1892} and Stoker~\cite{ref-stoker1957} and their
symmetric variants admit closed-form solutions; this idealization is
not a claim of physical generality but the precise condition under
which the solver can be verified end-to-end against exact mathematics.
Every more elaborate one-dimensional SWE solver, including those
incorporating bed slope, Manning friction, or lateral source terms,
must reduce to~\eqref{eq:sv_conservative} in the source-free limit and
must reproduce the elementary wave structures examined here; the
present library is intended to serve as a reference implementation
against which such extensions can be benchmarked.

Partition the computational domain $x \in [0, L]$ into $N$ cells of
uniform width $\Delta x = L/N$, with cell centers
$x_{j} = (j - \tfrac{1}{2})\,\Delta x$ for $j = 1, \ldots, N$. Cell
averages of the state vector~\eqref{eq:sv_state_flux} at the discrete
time $t^{n}$ are denoted
\begin{equation}
\mathbf{U}_{j}^{n} \;\approx\;
\frac{1}{\Delta x}\!\int_{x_{j} - \Delta x/2}^{x_{j} + \Delta x/2}\!
\mathbf{U}(x, t^{n})\,\mathrm{d}x.
\label{eq:cell_average}
\end{equation}
Integrating~\eqref{eq:sv_conservative} over the $j$-th cell and over
the time interval $[t^{n}, t^{n+1}]$, $\Delta t \equiv t^{n+1} - t^{n}$,
and applying the divergence theorem yields the exact finite-volume
update~\cite{ref-leveque2002}
\begin{equation}
\mathbf{U}_{j}^{n+1} \;=\;
\mathbf{U}_{j}^{n}
\;-\;
\frac{\Delta t}{\Delta x}\,
\bigl[\,\mathbf{F}_{j+1/2} - \mathbf{F}_{j-1/2}\,\bigr],
\label{eq:fv_update}
\end{equation}
in which $\mathbf{F}_{j \pm 1/2}$ are time-averaged numerical fluxes
across the interfaces $x = x_{j} \pm \Delta x/2$. The discretization
\eqref{eq:fv_update} is conservative by construction: summing over
$j$ telescopes the interior fluxes, so the discrete mass and momentum
inside the domain change only through the boundary fluxes
$\mathbf{F}_{1/2}$ and $\mathbf{F}_{N+1/2}$. The numerical scheme is
specified by the reconstruction that supplies left and right interface
states $\mathbf{U}_{j+1/2}^{L}, \mathbf{U}_{j+1/2}^{R}$ from the cell
averages, and by the Riemann-solver map
$\mathbf{F}_{j+1/2} = \widehat{\mathbf{F}}(\mathbf{U}_{j+1/2}^{L},
\mathbf{U}_{j+1/2}^{R})$.

Second-order spatial accuracy in smooth regions, with the local order
reduced to first order at extrema and discontinuities, is recovered by
the MUSCL linear reconstruction~\cite{ref-vanleer1979} with the minmod
slope limiter. For each conserved component $U_{j}^{(k)}$ with
$k \in \{1, 2\}$, the limited slope is
\begin{equation}
s_{j}^{(k)} \;=\;
\mathrm{minmod}\!\left(
\frac{U_{j}^{(k)} - U_{j-1}^{(k)}}{\Delta x},\;
\frac{U_{j+1}^{(k)} - U_{j}^{(k)}}{\Delta x}
\right),
\label{eq:slope_minmod}
\end{equation}
where
\begin{equation}
\mathrm{minmod}(a, b) \;=\;
\begin{cases}
\mathrm{sgn}(a)\,\min\!\bigl(|a|, |b|\bigr), & a\,b > 0,\\[2pt]
0, & a\,b \le 0,
\end{cases}
\label{eq:minmod_def}
\end{equation}
which is the most diffusive of the standard symmetric total-variation
diminishing limiters and is chosen for its robustness near shocks and
the dry front. At cells adjacent to the lateral domain boundaries the
slopes are evaluated by one-sided differences without limiting; the
attendant first-order accuracy in two cells is innocuous when
transmissive boundary data are imposed (below). The interface states
are
\begin{equation}
\mathbf{U}_{j+1/2}^{L} \;=\; \mathbf{U}_{j} + \tfrac{1}{2}\,\Delta x\,
\mathbf{s}_{j},
\qquad
\mathbf{U}_{j+1/2}^{R} \;=\; \mathbf{U}_{j+1} - \tfrac{1}{2}\,\Delta x\,
\mathbf{s}_{j+1},
\label{eq:muscl}
\end{equation}
and the reconstructed depth components are floored at
$\varepsilon_{\mathrm{rec}} = 10^{-8}$~m to ensure that the Riemann
solver downstream of~\eqref{eq:muscl} operates on strictly positive
depths.

The interface flux is computed by the HLLC three-wave approximate
Riemann solver of Toro, Spruce, and Speares~\cite{ref-toro1994},
specialized to the SWE. For the one-dimensional system the middle
contact wave of HLLC is mathematically degenerate, since
\eqref{eq:sv_conservative} carries no advected tracer; the HLLC star
states nonetheless give a well-conditioned algorithmic form that
generalizes cleanly to tracer-extended and two-dimensional SWE solvers,
and we retain this form here for numerical and pedagogical reasons.
Given interface states $\mathbf{U}^{L}, \mathbf{U}^{R}$ with primitive
variables $u^{L,R} = q^{L,R}/h^{L,R}$ and celerities
$c^{L,R} = \sqrt{g\,h^{L,R}}$, the Roe-averaged intermediate
state~\cite{ref-einfeldt1988} is
\begin{equation}
\tilde{u} \;=\;
\frac{\sqrt{h^{L}}\,u^{L} + \sqrt{h^{R}}\,u^{R}}
     {\sqrt{h^{L}} + \sqrt{h^{R}}},
\qquad
\tilde{c} \;=\;
\sqrt{\tfrac{1}{2}\,g\,(h^{L} + h^{R})},
\label{eq:roe_average}
\end{equation}
and the outer wave-speed estimates follow the Einfeldt
prescription~\cite{ref-einfeldt1988},
\begin{equation}
S_{L} \;=\; \min\!\bigl(u^{L} - c^{L},\; \tilde{u} - \tilde{c}\bigr),
\qquad
S_{R} \;=\; \max\!\bigl(u^{R} + c^{R},\; \tilde{u} + \tilde{c}\bigr).
\label{eq:wave_speeds}
\end{equation}
The middle (contact) wave speed is determined by enforcing continuity
of normal velocity across the star region, giving
\begin{equation}
S_{\star} \;=\;
\frac{S_{L}\,h^{R}\,(u^{R} - S_{R})
    - S_{R}\,h^{L}\,(u^{L} - S_{L})}
     {h^{R}\,(u^{R} - S_{R})
    - h^{L}\,(u^{L} - S_{L})},
\label{eq:contact_speed}
\end{equation}
with a numerically conditioned fallback to the canonical
HLL flux~\cite{ref-hartenlaxleer1983} whenever the denominator
of~\eqref{eq:contact_speed} drops below $10^{-14}$ in absolute value.
The star states across the left and right outer waves are determined
by the Rankine--Hugoniot jump conditions across $S_{L,R}$ combined with
the contact-continuity condition $u^{\star,L} = u^{\star,R} = S_{\star}$,
yielding
\begin{equation}
h^{\star,L} \;=\; h^{L}\,\frac{S_{L} - u^{L}}{S_{L} - S_{\star}},
\qquad
h^{\star,R} \;=\; h^{R}\,\frac{S_{R} - u^{R}}{S_{R} - S_{\star}},
\qquad
q^{\star,L,R} \;=\; h^{\star,L,R}\,S_{\star}.
\label{eq:star_state}
\end{equation}
The HLLC flux assembled across the four signal regions is
\begin{equation}
\widehat{\mathbf{F}}^{\mathrm{HLLC}} \;=\;
\begin{cases}
\mathbf{F}^{L}, & 0 \le S_{L},\\[2pt]
\mathbf{F}^{L} + S_{L}\,\bigl(\mathbf{U}^{\star,L} - \mathbf{U}^{L}\bigr),
& S_{L} < 0 \le S_{\star},\\[2pt]
\mathbf{F}^{R} + S_{R}\,\bigl(\mathbf{U}^{\star,R} - \mathbf{U}^{R}\bigr),
& S_{\star} < 0 \le S_{R},\\[2pt]
\mathbf{F}^{R}, & S_{R} < 0,
\end{cases}
\label{eq:hllc_flux}
\end{equation}
where $\mathbf{F}^{L,R} \equiv \mathbf{F}(\mathbf{U}^{L,R})$ are the
physical fluxes evaluated from~\eqref{eq:sv_state_flux} at the
reconstructed states. The flux~\eqref{eq:hllc_flux} is consistent,
conservative, and entropy-satisfying in the wet-state
regime~\cite{ref-toro2001}.

Time integration of the semi-discrete system implied
by~\eqref{eq:fv_update}, written compactly as $\mathrm{d}
\mathbf{U}_{j}/\mathrm{d}t = \mathcal{L}(\mathbf{U})_{j}$ with
$\mathcal{L}(\mathbf{U})_{j} = -(\mathbf{F}_{j+1/2} -
\mathbf{F}_{j-1/2})/\Delta x$, is performed by the two-stage
second-order SSP-RK method of Shu and
Osher~\cite{ref-shuosher1988} in convex-combination
form~\cite{ref-gottlieb2001},
\begin{equation}
\mathbf{U}^{(1)} \;=\;
\mathbf{U}^{n} + \Delta t\,\mathcal{L}(\mathbf{U}^{n}),
\qquad
\mathbf{U}^{n+1} \;=\;
\tfrac{1}{2}\,\mathbf{U}^{n}
+ \tfrac{1}{2}\,\mathbf{U}^{(1)}
+ \tfrac{1}{2}\,\Delta t\,\mathcal{L}(\mathbf{U}^{(1)}).
\label{eq:ssprk2}
\end{equation}
The scheme~\eqref{eq:ssprk2} writes the updated state as a convex
combination of two forward-Euler substeps, so any monotonicity or
total-variation-diminishing property established for the forward-Euler
step under a time-step constraint $\Delta t \le \Delta t_{\mathrm{FE}}$
is inherited by~\eqref{eq:ssprk2} under the same
constraint~\cite{ref-gottlieb2001}, i.e., the strong-stability-preserving
Courant--Friedrichs--Lewy (CFL) coefficient is unity. After each substage
of~\eqref{eq:ssprk2}, the boundary conditions and the positivity
correction described below are applied to the intermediate state, so
that the second-stage right-hand side $\mathcal{L}(\mathbf{U}^{(1)})$
is always evaluated on an admissible (positive-depth) field.

The time step is selected adaptively at every integration step by
the CFL condition
\begin{equation}
\Delta t \;=\;
\mathrm{CFL}\,\frac{\Delta x}{\displaystyle
\max_{1 \le j \le N}\bigl(|u_{j}^{n}| + c_{j}^{n}\bigr)},
\qquad
c_{j}^{n} \;=\; \sqrt{g\,h_{j}^{n}},
\label{eq:cfl}
\end{equation}
where the maximum signal speed is determined by the
eigenvalues~\eqref{eq:sv_eigenvalues} of the flux Jacobian
\eqref{eq:sv_jacobian}, and the target CFL number is set to $0.9$ in
all reported runs. The velocity $u_{j}^{n} = q_{j}^{n}/h_{j}^{n}$ is
computed with a guard that returns zero whenever $h_{j}^{n}$ falls
below the positivity floor.

Lateral boundary conditions (BC) are imposed by zero-gradient (transmissive)
extrapolation,
\begin{equation}
h_{0} \;=\; h_{1},
\qquad h_{N+1} \;=\; h_{N},
\qquad q_{0} \;=\; q_{1},
\qquad q_{N+1} \;=\; q_{N},
\label{eq:bc_transmissive}
\end{equation}
which is a first-order non-reflecting approximation for waves
propagating outward through the edges of the
domain~\cite{ref-toro2001}. Strict non-reflection would require
characteristic decomposition at the boundary;
\eqref{eq:bc_transmissive} is sufficient for the present idealized
benchmarks, in which the domain length $L$ is chosen large enough that
the wave structure of interest remains interior over the simulation
horizon.

Degeneracy of the system~\eqref{eq:sv_conservative} at the dry state,
where the two eigenvalues~\eqref{eq:sv_eigenvalues} coalesce as
$h \to 0$, is handled by a two-tier positivity strategy. After each
substage of~\eqref{eq:ssprk2} and after the application
of~\eqref{eq:bc_transmissive}, the cell-centered depth is clamped above
the floor $\varepsilon_{h} = 10^{-8}$~m,
\begin{equation}
(h_{j}, q_{j}) \;\longleftarrow\;
\begin{cases}
(h_{j},\; q_{j}), & h_{j} \ge \varepsilon_{h},\\[2pt]
(\varepsilon_{h},\; 0), & h_{j} < \varepsilon_{h}.
\end{cases}
\label{eq:positivity_fix}
\end{equation}
Equation~\eqref{eq:positivity_fix} maintains strictly positive depth
in every cell while zeroing the residual discharge in newly dried
cells, preserving the conservative form
of~\eqref{eq:fv_update} up to the controlled deviation introduced
by the floor. The MUSCL face reconstruction~\eqref{eq:muscl} is
independently floored at the same value $\varepsilon_{h}$ before the
Riemann solve. A second, larger threshold
$H_{\mathrm{DRY}} = 10^{-2}$~m governs the post-processing diagnostics
that depend on $u = q/h$, including the maximum Froude number and the
wet-cell velocity error norm; this separation between the solver-level
positivity floor and the physics-level wet-cell threshold prevents the
diagnostic from being contaminated by cells where $u$ is dominated by
the floor.

The three primary discrete invariants tracked at every time step are
\begin{equation}
M^{n} \;=\; \Delta x \sum_{j=1}^{N} h_{j}^{n},
\qquad
P^{n} \;=\; \Delta x \sum_{j=1}^{N} q_{j}^{n},
\qquad
E^{n} \;=\; \Delta x \sum_{j=1}^{N}
\Bigl(\tfrac{1}{2}\,u_{j}^{n}\,q_{j}^{n}
\;+\; \tfrac{1}{2}\,g\,(h_{j}^{n})^{2}\Bigr),
\label{eq:discrete_invariants}
\end{equation}
which correspond to total water mass, total momentum, and total
mechanical (kinetic plus potential) energy per unit channel width,
integrated over the computational domain. The conservative
discretization~\eqref{eq:fv_update} guarantees that $M^{n}$ and
$P^{n}$ are exactly conserved up to boundary fluxes; with the
transmissive boundaries~\eqref{eq:bc_transmissive}, departures of
$M^{n}$ and $P^{n}$ from their initial values diagnose physical efflux
of mass and momentum through the lateral edges and not solver error,
provided $\Delta t \le \Delta t_{\mathrm{CFL}}$
in~\eqref{eq:cfl}. The energy
$E^{n}$ decreases monotonically across shocks in accordance with the
Lax entropy condition for the SWE, a property of the HLLC
flux~\eqref{eq:hllc_flux} in the entropy-satisfying
regime~\cite{ref-toro2001}; we record the cumulative energy
dissipation as a diagnostic of shock-capturing fidelity. The
maximum Froude number
$\mathrm{Fr}^{n} = \max_{j:\,h_{j}^{n} > H_{\mathrm{DRY}}}
|u_{j}^{n}|/c_{j}^{n}$ provides an instantaneous indicator of the
local flow regime and is used to confirm subcritical or supercritical
behavior in each test case.

The solver is distributed under the MIT license as the Python package
\texttt{amerta}, with source code at
\url{https://github.com/sandyherho/amerta} and a release on the Python
Package Index at \url{https://pypi.org/project/amerta/}, implemented in
pure Python with selective JIT compilation. The performance-critical
inner kernels, namely the minmod limiter~\eqref{eq:minmod_def}, the
MUSCL reconstruction~\eqref{eq:muscl}, the HLLC
flux~\eqref{eq:hllc_flux}, and the SSP-RK2
stage~\eqref{eq:ssprk2}, are compiled to LLVM-optimized machine code
by Numba~\cite{ref-numba} using the \texttt{@njit(cache=True)}
decorator; the loops over cells and interfaces are parallelized over
CPU threads through \texttt{@njit(parallel=True)} together with
\texttt{prange}. Array primitives, the analytical-solution evaluators,
and the spectral diagnostics rely on NumPy~\cite{ref-numpy}, and the
bracketed bisection root finder \texttt{scipy.optimize.brentq} from
SciPy~\cite{ref-scipy} solves the Rankine--Hugoniot star-state
nonlinear equation for the Stoker and double-shock analytical
benchmarks. Spatiotemporal output is written to CF-1.8-compliant
NetCDF-4 files via the \texttt{netCDF4} Python
interface~\cite{ref-netcdf4python} and to comma-separated-value
tabular summaries via Pandas~\cite{ref-pandas}. Static figures and
animated GIFs are produced by Matplotlib~\cite{ref-matplotlib} using
the Pillow writer backend~\cite{ref-pillow}, and progress reporting
during integration uses tqdm~\cite{ref-tqdm}.

The library is distributed with four canonical Riemann-problem
configurations that, taken together, exhaust the elementary wave types
admissible by~\eqref{eq:sv_conservative} in the absence of source
terms: a rarefaction joined to a shock through a constant intermediate
state, a single rarefaction terminating at a dry front, two diverging
rarefactions, and two converging shocks. All four configurations share
the dam-break initial datum
\begin{equation}
\mathbf{U}(x, 0) \;=\;
\begin{cases}
(h_{L},\; h_{L}\,u_{L})^{T}, & x < L/2,\\[2pt]
(h_{R},\; h_{R}\,u_{R})^{T}, & x \ge L/2,
\end{cases}
\label{eq:ic_dambreak}
\end{equation}
on the domain $x \in [0, L]$ with gravitational acceleration
$g = 9.81$~m\,s$^{-2}$ and transmissive
boundaries~\eqref{eq:bc_transmissive}.

The first configuration (Stoker) sets $h_{L} = 10$~m, $h_{R} = 2$~m,
$u_{L} = u_{R} = 0$, $L = 2000$~m, $t_{\mathrm{final}} = 80$~s, and
$N = 500$. The exact solution~\cite{ref-stoker1957} consists of a
left-going rarefaction fan connecting the upstream rest state to an
intermediate plateau $(h_{\star}, u_{\star})$ determined by the
Riemann invariants of the 1-family, followed by a right-going shock
advancing into the lower downstream water with speed $S$ obtained
from the Rankine--Hugoniot jump condition. The intermediate depth
$h_{\star}$ is found by Newton iteration on the nonlinear matching
condition $u_{L} + 2(c_{L} - c_{\star}) = u_{R} + (h_{\star} - h_{R})
\sqrt{g(h_{\star} + h_{R})/(2 h_{\star} h_{R})}$. Physically, this is
the idealized prototype of a partial dam breach into a non-empty
downstream channel; the constant-intermediate plateau between the
rarefaction tail and the shock is a generic feature of any two-state
hyperbolic Riemann fan and reappears in the wet portion of every more
realistic SWE dam-break scenario.

The second configuration (Ritter) sets $h_{L} = 10$~m,
$h_{R} = 10^{-3}$~m, $u_{L} = u_{R} = 0$, $L = 2000$~m,
$t_{\mathrm{final}} = 40$~s, and $N = 500$, and is the wet-dry dam
break first analyzed in closed form by Ritter~\cite{ref-ritter1892}.
The right state is assigned the small positive value
$h_{R} = 10^{-3}$~m rather than zero in order to match the solver
positivity floor~\eqref{eq:positivity_fix}; the exact solution against
which the numerical result is compared is the genuine dry-bed Ritter
rarefaction in the limit $h_{R} \to 0^{+}$, with depth profile
$h(x, t) = (2 c_{L} - \xi)^{2}/(9 g)$ and velocity
$u(x, t) = (2/3)(\xi + c_{L})$ for $\xi \equiv (x - L/2)/t$ in the
fan region $-c_{L} \le \xi \le 2 c_{L}$, vanishing identically ahead
of the dry front $\xi > 2 c_{L}$. The dry front advances at
$2 c_{L} = 2\sqrt{g\,h_{L}} \approx 19.81$~m\,s$^{-1}$ relative to the
dam. Physically, this corresponds to total dam failure onto an
initially dry channel or to flood inundation across previously dry
terrain; the wave structure tests the solver's capacity to maintain
positivity at a moving free boundary where the
system~\eqref{eq:sv_conservative} loses strict hyperbolicity.

The third configuration (symmetric double rarefaction) sets
$h_{L} = h_{R} = 5$~m, $u_{L} = -3$~m\,s$^{-1}$,
$u_{R} = +3$~m\,s$^{-1}$, $L = 2000$~m, $t_{\mathrm{final}} = 80$~s,
and $N = 1000$. The diverging-flow initial condition develops two
symmetric rarefaction fans propagating outward from the central
discontinuity, with a central drawdown state of depth
$h_{\star} = c_{\star}^{2}/g$ where
$c_{\star} = (u_{L} + 2 c_{L})/2$ follows from intersection of the
1-family and 2-family Riemann invariants at the symmetric center;
for the present parameters $h_{\star} \approx 3.08$~m, well within
the wet-star regime $|u_{L,R}| < 2 c_{L,R} \approx 14$~m\,s$^{-1}$
that separates this case from cavitation. Physically, the
configuration is an idealization of bidirectional drawdown around a
withdrawn obstruction, of the transient at the opening of a tidal or
sluice gate, or of the central section of an extended flood-recession
profile. The absence of compressive waves makes this case a clean
test of the solver's behavior in genuinely rarefactive regimes and a
sensitive probe of artificial steepening from the slope
limiter~\eqref{eq:slope_minmod}.

The fourth configuration (symmetric double shock) sets
$h_{L} = h_{R} = 3$~m, $u_{L} = +3$~m\,s$^{-1}$,
$u_{R} = -3$~m\,s$^{-1}$, $L = 2000$~m, $t_{\mathrm{final}} = 80$~s,
and $N = 500$, and is the time-reverse symmetric counterpart of the
third case. Two compressive waves converge to produce a single
elevated star state $h_{\star} > h_{R}$ flanked by two symmetric
shocks of speed $\pm S$ obtained from $S = h_{R}\,u_{L}/(h_{R} -
h_{\star})$, with $h_{\star}$ determined by Newton iteration on the
Rankine--Hugoniot condition $u_{L} = (h_{\star} - h_{R})
\sqrt{g(h_{\star} + h_{R})/(2 h_{\star} h_{R})}$. Physically, this
represents the convergence of two opposing flood waves, the merging
of two tidal bores, or the central section of a flow-on-flow
collision; it provides the most stringent test of the HLLC
shock-capturing accuracy~\eqref{eq:hllc_flux} and of the
entropy-consistent dissipation embedded in the discrete energy
$E^{n}$, since the entire wave structure consists of discontinuities
without smooth components.

The four configurations described above are stored as plain-text
\texttt{key = value} files in the \texttt{configs/} directory of the
source distribution and are loaded by the \texttt{ConfigManager}
class. The same interface accepts user-supplied configurations, so
that any wet or dry Riemann problem of the form~\eqref{eq:ic_dambreak}
on a horizontal frictionless bed can be simulated by specifying the
parameters $(h_{L}, h_{R}, u_{L}, u_{R}, L, t_{\mathrm{final}}, N,
\mathrm{CFL}, g)$ without modification to the solver source. Solver
verification against the corresponding closed-form analytical
solution is automatic for any user-defined case whose
\texttt{case\_type} matches one of the four canonical labels; for
other initial data, the conservation diagnostics
\eqref{eq:discrete_invariants} and the maximum-Froude indicator
remain available and provide solver-independent quality checks.

We emphasize that the present model and the four embedded cases are
deliberately idealized. Bed slope $S_{0}$, bottom friction (Manning,
Ch{\'e}zy, or Darcy--Weisbach closures), lateral mass and momentum
source terms (rainfall, infiltration, tributary inflow), channel
cross-section variability, transverse flow components, and viscous
or turbulent dissipation are all set to zero
in~\eqref{eq:sv_conservative} and in the discretization
\eqref{eq:fv_update}--\eqref{eq:hllc_flux}. The four embedded cases
are exactly those for which closed-form analytical solutions
of~\eqref{eq:sv_conservative} are available for end-to-end solver
verification. The motivation for this restriction is mathematical
traceability rather than physical realism. The canonical Riemann
problems~\cite{ref-ritter1892,ref-stoker1957} are the only one-dimensional
SWE configurations for which the wave structure can be written in
closed form, and any deviation between numerical and analytical
solutions in these cases can be assigned unambiguously to the
solver. Operationally relevant flood and dam-break models incorporate
source terms that destroy this analytical tractability and require
manufactured-solution or grid-refinement studies for verification;
such models still must reproduce the canonical Riemann-fan structures
in their source-free limit, and the \texttt{amerta} library provides
a reference implementation against which extensions incorporating
bed slope, Manning friction, channel-geometry variability, or
two-dimensional generalization can be benchmarked.

\subsection{Data Analysis}

The numerical outputs produced by the \texttt{amerta} solver, written
as CF-1.8 NetCDF-4 files containing the complete spatiotemporal
trajectory $\bigl(h_{j}^{n},\, u_{j}^{n},\, q_{j}^{n}\bigr)$ at every
solver timestep together with the discrete
invariants~\eqref{eq:discrete_invariants} and the stored
analytical-comparison error norms, are post-processed by an
independent suite of six Python scripts that we describe in this
subsection. The post-processing depends only on
NumPy~\cite{ref-numpy}, SciPy~\cite{ref-scipy}, and
Matplotlib~\cite{ref-matplotlib}; no information beyond what is
archived in the NetCDF files is required, so the diagnostics are
fully reproducible from the solver outputs alone. Each script
isolates one aspect of solution quality, namely qualitative
wave-structure recognition, pointwise agreement with closed-form
solutions, self-similar collapse of the Riemann fan, time evolution
of integral error norms, boundary-flux-corrected conservation
balance, and confinement of the discrete state to the analytical
wave curves in the conserved-variable plane. Taken together, the
six diagnostics constitute a verification audit of the solver under
the four canonical configurations introduced above, and apply equally to
any user-defined Riemann problem of the form~\eqref{eq:ic_dambreak}
supported by the library.

\subsubsection{Space-time visualization of the depth field}

The first diagnostic renders the spatiotemporal depth field
$h_{j}^{n}$ produced by~\eqref{eq:fv_update} as a three-dimensional
surface over $(x, t, h)$ for each of the four test cases. To
suppress rasterization cost and visual aliasing of fine-scale
numerical features without affecting macroscopic wave structure,
the field is subsampled on a fixed grid of
$N_{x}^{\mathrm{plot}} = 200$ spatial nodes and
$N_{t}^{\mathrm{plot}} = 80$ temporal snapshots, selected by
linear index spacing across the full stored trajectory. The four
panels share a common viridis colormap whose range
$\bigl[\min_{c, j, n} h_{j}^{n, c},\;
       \max_{c, j, n} h_{j}^{n, c}\bigr]$
is computed globally over all four cases
$c \in \{1, 2, 3, 4\}$, so that relative depth magnitudes are
directly comparable across panels, and a fixed viewing geometry
(elevation $28^{\circ}$, azimuth $-60^{\circ}$) is used for all
four. Although qualitative, this diagnostic functions as the
first-pass topological check that the expected Riemann-fan
structure, namely the rarefaction-shock pair of Stoker, the single
rarefaction with dry front of Ritter, two diverging rarefactions,
and two converging shocks, is present in the numerical output
before quantitative tests are applied. The rendering uses the
\texttt{mpl\_toolkits.mplot3d} module of
Matplotlib~\cite{ref-matplotlib}.

\subsubsection{Final-time validation against analytical Riemann solutions}

The second diagnostic overlays the numerical and analytical depth
profiles at the final time $t = t_{\mathrm{final}}$, with the
analytical profile $h_{\mathrm{an}}(x, t_{\mathrm{final}})$
constructed from the closed-form Riemann solutions described
above. The pointwise depth error is
\begin{equation}
e_{j}(t_{\mathrm{final}}) \;\equiv\;
h_{j}^{N_{t}} \;-\; h_{\mathrm{an}}(x_{j}, t_{\mathrm{final}}),
\label{eq:pointwise_error}
\end{equation}
and the integral norms $L^{1}(h)(t_{\mathrm{final}})$ and
$L^{2}(h)(t_{\mathrm{final}})$ stored in the NetCDF file are
reported directly. The empirical distribution of
$\{|e_{j}(t_{\mathrm{final}})|\}_{j=1}^{N}$ is summarized through
the quantiles $Q_{p}\bigl[|e|\bigr]$ at percentiles
$p \in \{50,\, 75,\, 90,\, 95,\, 99,\, 99.9\}$, evaluated by the
linear-interpolation estimator implemented in
\texttt{numpy.percentile}~\cite{ref-numpy}.

The motivation for reporting a full quantile sweep rather than a
single summary statistic such as the root-mean-square error is the
nonuniform spatial distribution of error in shock-capturing
schemes~\cite{ref-harten1983}: the
limiter~\eqref{eq:slope_minmod} reduces to first order in a small
number of cells adjacent to a shock front or at the corners of a
rarefaction fan, where the local truncation error is concentrated.
The lower percentiles $p \in \{50,\, 75,\, 90\}$ characterize the
bulk accuracy in smooth regions, while the tail percentiles
$p \in \{99,\, 99.9\}$ isolate the localized error contribution
from the dissipative cells. The spread between the median and the
$99.9$-th percentile is therefore a direct empirical measure of
how localized the scheme's truncation error is around
discontinuities.

\subsubsection{Self-similarity collapse of the Riemann fan}

The third diagnostic exploits the scaling invariance of the
Riemann problem and does not require access to the analytical
solution. Equations~\eqref{eq:sv_conservative} together with the
dam-break initial datum~\eqref{eq:ic_dambreak} are invariant under
$(x, t) \mapsto (\alpha\,x,\, \alpha\,t)$ for any $\alpha > 0$,
and consequently any classical or entropy weak
solution~\cite{ref-lax1957} depends on $x$ and $t$ only through
the similarity variable
\begin{equation}
\xi \;\equiv\; \frac{x - x_{\mathrm{dam}}}{t},
\qquad x_{\mathrm{dam}} \;=\; L/2,
\label{eq:similarity_var}
\end{equation}
that is, $h(x, t) = H(\xi)$ and $u(x, t) = U(\xi)$ everywhere away
from $t = 0$. For each of the four cases the diagnostic
reparameterizes the numerical depth as $H_{\mathrm{num}}(\xi)$ for
$N_{\mathrm{ov}} = 12$ time snapshots taken at evenly spaced
indices within the window $t > 0.05\,t_{\mathrm{final}}$, where
the initial $5\%$ of the simulation is excluded to remove
transient adjustment from the discontinuous initial
datum~\eqref{eq:ic_dambreak}, and overlays these snapshots in a
single $(\xi, h)$ panel together with the analytical curve
$H_{\mathrm{an}}(\xi)$ extracted from the last stored snapshot.

A scalar similarity-quality index is constructed by partitioning
the range of $\xi$ values present in the data into $N_{b} = 200$
uniform bins $B_{b}$, $b = 1, \ldots, N_{b}$. Within each bin we
evaluate the temporal-scatter standard deviation
\begin{equation}
\sigma_{b} \;\equiv\;
\mathrm{std}\!\Bigl\{\,h_{j}^{n}
\;:\; \xi(x_{j}, t^{n}) \in B_{b},\;
t^{n} > 0.05\,t_{\mathrm{final}}\,\Bigr\},
\label{eq:sigma_bin}
\end{equation}
defined whenever the bin contains at least two samples and
treated as missing otherwise; bins for which $\sigma_{b}$ is
undefined are excluded from the average. The dimensionless
similarity quality is
\begin{equation}
\mathcal{Q}_{\mathrm{sim}} \;\equiv\;
1 \;-\; \frac{\bigl\langle \sigma_{b} \bigr\rangle_{b}}
              {\bigl\langle h \bigr\rangle},
\qquad
\bigl\langle h \bigr\rangle \;=\;
\frac{1}{N_{t}\,N}\,
\sum_{n=0}^{N_{t}-1}\sum_{j=1}^{N} h_{j}^{n},
\label{eq:Qsim}
\end{equation}
where $\langle \sigma_{b} \rangle_{b}$ denotes the arithmetic mean
over the bins where $\sigma_{b}$ is defined. For an exactly
self-similar solution all snapshots collapse onto a single curve
in $(\xi, h)$ space, $\sigma_{b} \equiv 0$ for every bin, and
$\mathcal{Q}_{\mathrm{sim}} = 1$; departures of
$\mathcal{Q}_{\mathrm{sim}}$ from unity quantify cumulative solver
drift off the similarity manifold, which can arise from temporal
truncation error that grows with $t$ or from inappropriate
boundary-induced contamination of the interior wave structure.
Because $\mathcal{Q}_{\mathrm{sim}}$ is constructed entirely from
the numerical trajectory, it applies as a solver-internal
consistency test to user-defined Riemann problems for which no
closed-form solution exists.

\subsubsection{Time evolution of integral error norms}

The fourth diagnostic extracts and plots, on a logarithmic
ordinate, the four time series of integral error norms written by
the solver to NetCDF at every timestep. For a generic discretely
sampled field $\Psi$ with numerical and analytical values
$\Psi_{j}^{n, \mathrm{num}}$ and $\Psi_{j}^{n, \mathrm{an}}$,
define the discrete $L^{p}$ norm
\begin{equation}
L^{p}(\Psi)(t^{n}) \;\equiv\;
\Biggl[\,\Delta x \,\sum_{j=1}^{N}
\bigl|\,\Psi_{j}^{n, \mathrm{num}}
       - \Psi_{j}^{n, \mathrm{an}}\,\bigr|^{p}\,\Biggr]^{1/p},
\qquad p \in \{1, 2\}.
\label{eq:Lp_norm}
\end{equation}
The four reported series are $L^{1}(h)$, $L^{2}(h)$, $L^{1}(q)$,
and $L^{1}(u_{\mathrm{wet}})$, where the wet-cell velocity error
norm restricts the sum in~\eqref{eq:Lp_norm} to the index set
\begin{equation}
\mathcal{W}^{n} \;\equiv\;
\Bigl\{\,j \in \{1, \ldots, N\} \;:\;
h_{j}^{n} > H_{\mathrm{DRY}} \;\;\text{and}\;\;
h_{\mathrm{an}}(x_{j}, t^{n}) > H_{\mathrm{DRY}}\,\Bigr\},
\label{eq:wet_set}
\end{equation}
with $H_{\mathrm{DRY}} = 10^{-2}$~m the diagnostic wet-cell
threshold introduced above.

The choice of metrics is dictated by the structure of the SWE
Riemann solutions and by the dry-state degeneracy of the
system~\eqref{eq:sv_conservative}. The depth norms $L^{1}(h)$ and
$L^{2}(h)$ are standard~\cite{ref-leveque2002}: $L^{1}$ measures
total volumetric displacement of the wave structure and is the
natural pairing with the conserved variable $h$, while $L^{2}$
weights localized errors at shocks more heavily. The discharge
norm $L^{1}(q)$ is used in place of $L^{1}(u)$ for momentum-error
assessment because $q = h\,u$ is the conserved variable and is
bounded near a dry front, whereas $u = q/h$ diverges as the solver
positivity floor~\eqref{eq:positivity_fix} is approached;
consequently $L^{1}(u)$ is dominated by floor-driven artifacts in
the Ritter problem, while $L^{1}(q)$ remains a faithful measure of
the momentum-flux discrepancy throughout the dry-bed evolution.
The supplementary wet-cell norm $L^{1}(u_{\mathrm{wet}})$ recovers
a kinematic velocity diagnostic on the subset of cells where
$u = q/h$ is physically meaningful on both the numerical and
analytical sides, and is reported alongside $L^{1}(q)$ for the
dry-bed cases. The logarithmic ordinate is used so that asymptotic
time behavior can be inspected across several decades.

\subsubsection{Boundary-flux-corrected conservation diagnostics}

Under the transmissive boundary
condition~\eqref{eq:bc_transmissive}, the discrete invariants
$M^{n}$ and $E^{n}$ defined in~\eqref{eq:discrete_invariants} are
not conserved in time because mass and energy fluxes across the
lateral boundaries are admitted by construction. Departures of
these quantities from their initial values therefore reflect both
genuine physical efflux and any solver-level conservation error,
and a meaningful diagnostic must explicitly subtract the boundary
contribution. We construct two such diagnostics.

The cumulative net mass inflow through the lateral boundaries is
approximated by the composite trapezoidal rule applied to the
boundary-cell discharges,
\begin{equation}
\Phi_{M}(t^{n}) \;\equiv\;
\int_{0}^{t^{n}}\!
\bigl[\,q(x_{1}, s) \;-\; q(x_{N}, s)\,\bigr]\,\mathrm{d}s
\;\approx\;
\sum_{k=0}^{n-1} \tfrac{1}{2}\,(t^{k+1} - t^{k})
\bigl[\,(q_{1}^{k+1} - q_{N}^{k+1})
\;+\; (q_{1}^{k} - q_{N}^{k})\,\bigr],
\label{eq:mass_inflow}
\end{equation}
in which the leftmost and rightmost cell-centered discharges
$q_{1}^{n}$ and $q_{N}^{n}$ serve as boundary-trace approximants
by virtue of the zero-gradient
extrapolation~\eqref{eq:bc_transmissive}. The
boundary-flux-corrected mass residual is then
\begin{equation}
r_{M}(t^{n}) \;\equiv\;
\bigl(M^{n} - M^{0}\bigr) \;-\; \Phi_{M}(t^{n}),
\label{eq:mass_residual}
\end{equation}
which vanishes to machine precision for an exactly conservative
discretization of~\eqref{eq:sv_mass}. Equation~\eqref{eq:mass_residual}
isolates solver-induced mass creation or destruction from physical
efflux through the open boundaries, and is the proper conservation
metric for the present transmissive-BC configuration.

The energy balance is constructed analogously. For smooth
solutions the SWE energy density
$\mathcal{E}(h, u) \equiv \tfrac{1}{2}\,u^{2}\,h + \tfrac{1}{2}\,g\,h^{2}$,
which is the integrand of $E^{n}$
in~\eqref{eq:discrete_invariants}, satisfies the additional
conservation law
$\partial_{t}\mathcal{E} + \partial_{x} F_{\mathcal{E}} = 0$ with
flux
\begin{equation}
F_{\mathcal{E}}(h, u) \;\equiv\;
u\,\Bigl(\tfrac{1}{2}\,u^{2}\,h \;+\; g\,h^{2}\Bigr)
\qquad [\mathrm{m}^{4}\,\mathrm{s}^{-3}],
\label{eq:energy_flux}
\end{equation}
obtained by combining the SWE mass and momentum
laws~\eqref{eq:sv_mass} and~\eqref{eq:sv_momentum}. The Lax
entropy-admissibility condition~\cite{ref-lax1957} for the SWE
requires that the weak solution satisfy
$\partial_{t}\mathcal{E} + \partial_{x} F_{\mathcal{E}} \le 0$
across shocks, with strict inequality at admissible
discontinuities; any consistent entropy-stable shock-capturing
scheme must reproduce this property at the discrete
level~\cite{ref-tadmor1987}. Integrating in space and time, the
cumulative net energy inflow is
\begin{equation}
\Phi_{E}(t^{n}) \;\equiv\;
\int_{0}^{t^{n}}\!
\Bigl[\,F_{\mathcal{E}}\bigl(h(x_{1}, s),\, u(x_{1}, s)\bigr)
\;-\; F_{\mathcal{E}}\bigl(h(x_{N}, s),\, u(x_{N}, s)\bigr)\,\Bigr]\,
\mathrm{d}s,
\label{eq:energy_inflow}
\end{equation}
again approximated by the composite trapezoidal rule applied to
the boundary-cell traces. The cumulative dissipation produced by
the discretization over the interval $[0,\, t^{n}]$ is
\begin{equation}
D(t^{n}) \;\equiv\;
\Phi_{E}(t^{n}) \;-\; \bigl(E^{n} - E^{0}\bigr)
\qquad [\mathrm{m}^{4}\,\mathrm{s}^{-2}],
\label{eq:dissipation}
\end{equation}
and the discrete entropy-admissibility condition that any
consistent shock-capturing scheme must satisfy is
\begin{equation}
D(t^{n}) \;\ge\; 0 \quad \text{for all } n \;\ge\; 0.
\label{eq:entropy_admissibility}
\end{equation}
We report $\min_{n} D(t^{n})$ as a sign check
on~\eqref{eq:entropy_admissibility} and $D(t_{\mathrm{final}})$
as a magnitude estimate of the cumulative shock dissipation
produced by the HLLC flux~\eqref{eq:hllc_flux} in each case.

Two further diagnostics are recorded. The maximum wet-cell Froude
number with a strict positivity threshold,
\begin{equation}
\mathrm{Fr}_{\mathrm{strict}}^{n} \;\equiv\;
\max_{j \,:\, h_{j}^{n} > H_{\mathrm{strict}}}
\frac{|u_{j}^{n}|}{c_{j}^{n}},
\qquad
H_{\mathrm{strict}} \;=\; 5\,H_{\mathrm{DRY}} \;=\; 0.05~\mathrm{m},
\label{eq:strict_froude}
\end{equation}
demarcates subcritical
($\mathrm{Fr}_{\mathrm{strict}}^{n} < 1$) from supercritical
($\mathrm{Fr}_{\mathrm{strict}}^{n} > 1$) flow without
contamination from cells whose depth merely exceeds the diagnostic
threshold $H_{\mathrm{DRY}}$. The total variation of the conserved
discharge across the cells,
\begin{equation}
\mathrm{TV}(q)(t^{n}) \;\equiv\;
\sum_{j=1}^{N-1} \bigl|\,q_{j+1}^{n} - q_{j}^{n}\,\bigr|,
\label{eq:TVq}
\end{equation}
provides the proper test of the TVD
property introduced by Harten~\cite{ref-harten1983}: under
MUSCL-minmod reconstruction~\eqref{eq:slope_minmod}--\eqref{eq:muscl}
combined with the SSP-RK2 time integration~\eqref{eq:ssprk2}, the
scheme is formally TVD for the scalar advection equation under the
strong-stability-preserving CFL condition~\cite{ref-gottlieb2001},
and the empirical growth
\begin{equation}
\Delta\mathrm{TV}(q)(t^{n}) \;\equiv\;
\mathrm{TV}(q)(t^{n}) \;-\; \mathrm{TV}(q)(t^{0})
\label{eq:dTVq}
\end{equation}
quantifies how closely the property is preserved by the
discretization of the nonlinear hyperbolic
system~\eqref{eq:sv_conservative}. We report $\mathrm{TV}(q)$
rather than $\mathrm{TV}(u)$ on the same grounds as for the error
norms reported above: $q$ is the conserved variable, while
$u = q/h$ is contaminated by the positivity floor near dry states.

\subsubsection{Phase-plane analysis against analytical wave curves}

The sixth diagnostic projects the final-time numerical solution
$\bigl\{(h_{j}^{N_{t}},\, u_{j}^{N_{t}})\bigr\}_{j=1}^{N}$ onto
the conserved-variable plane $(h, u) \in \mathbb{R}^{2}_{+}$ and
overlays the analytical wave curves of the Riemann fan, on which
the entropy solution lies exactly~\cite{ref-lax1957}. The wave
curves connect the left state $(h_{L}, u_{L})$ and the right state
$(h_{R}, u_{R})$ through the intermediate (star) state
$(h_{\star}, u_{\star})$, and are determined by the eigenpair
structure~\eqref{eq:sv_eigenvalues}--\eqref{eq:sv_eigenvectors}.
The Riemann invariants of the SWE,
\begin{equation}
\mathcal{R}_{1}(\mathbf{U}) \;=\; u \;+\; 2\,c,
\qquad
\mathcal{R}_{2}(\mathbf{U}) \;=\; u \;-\; 2\,c,
\qquad
c \;=\; \sqrt{g\,h},
\label{eq:riemann_invariants}
\end{equation}
are constant along the right-going characteristic family
$\mathrm{d}x/\mathrm{d}t = \lambda_{+}$ and the left-going family
$\mathrm{d}x/\mathrm{d}t = \lambda_{-}$ of the
eigenvalues~\eqref{eq:sv_eigenvalues}, respectively. Constancy of
$\mathcal{R}_{1}$ across a 1-rarefaction emanating from
$(h_{L}, u_{L})$ yields the parametric curve in the phase plane
\begin{equation}
u(h) \;=\; u_{L} \;+\; 2\,\bigl(\sqrt{g\,h_{L}} - \sqrt{g\,h}\bigr),
\qquad 0 \;<\; h \;\le\; h_{L},
\label{eq:rarefaction_1}
\end{equation}
and analogously a 2-rarefaction emanating from $(h_{R}, u_{R})$
obeys
\begin{equation}
u(h) \;=\; u_{R} \;-\; 2\,\bigl(\sqrt{g\,h_{R}} - \sqrt{g\,h}\bigr),
\qquad 0 \;<\; h \;\le\; h_{R}.
\label{eq:rarefaction_2}
\end{equation}
For shocks, the Rankine--Hugoniot jump conditions combined with
the Lax admissibility criterion~\cite{ref-lax1957} give the
1-shock and 2-shock curves
\begin{equation}
u(h) \;=\; u_{L} \;-\; (h - h_{L})\,
\sqrt{\dfrac{g\,(h + h_{L})}{2\,h\,h_{L}}},
\qquad h \;>\; h_{L},
\label{eq:shock_1}
\end{equation}
\begin{equation}
u(h) \;=\; u_{R} \;+\; (h - h_{R})\,
\sqrt{\dfrac{g\,(h + h_{R})}{2\,h\,h_{R}}},
\qquad h \;>\; h_{R}.
\label{eq:shock_2}
\end{equation}
The star state $(h_{\star}, u_{\star})$ is the unique intersection
of the wave curves emanating from $(h_{L}, u_{L})$ and
$(h_{R}, u_{R})$ that respects the wave-type sequence of each
case~\cite{ref-toro2001}.

For the Stoker problem,~\eqref{eq:rarefaction_1}
intersects~\eqref{eq:shock_2}, and the matching condition
\begin{equation}
u_{L} \;+\; 2\,\bigl(\sqrt{g\,h_{L}} - \sqrt{g\,h_{\star}}\bigr)
\;=\;
u_{R} \;+\; (h_{\star} - h_{R})\,
\sqrt{\dfrac{g\,(h_{\star} + h_{R})}{2\,h_{\star}\,h_{R}}}
\label{eq:stoker_star_match}
\end{equation}
is solved for $h_{\star}$ on the bracket $(h_{R},\, h_{L})$ by
Brent's hybrid bisection-secant-inverse-quadratic
algorithm~\cite{ref-brent1971} as implemented in
\texttt{scipy.optimize.brentq}~\cite{ref-scipy}; $u_{\star}$ then
follows from~\eqref{eq:rarefaction_1}. For the symmetric double
rarefaction with $h_{L} = h_{R} \equiv h_{0}$ and
$c_{L} = c_{R} \equiv c_{0}$, intersection
of~\eqref{eq:rarefaction_1} with~\eqref{eq:rarefaction_2} gives
the closed-form star state
\begin{equation}
c_{\star} \;=\; c_{0} \;+\; \tfrac{1}{4}\,(u_{L} - u_{R}),
\qquad
h_{\star} \;=\; \max\!\bigl(c_{\star}^{2}/g,\; 0\bigr),
\qquad
u_{\star} \;=\; 0,
\label{eq:dr_star}
\end{equation}
in which the symmetry $u_{L} = -u_{R}$ implies $u_{\star} = 0$
exactly. For the symmetric double shock, intersection
of~\eqref{eq:shock_1} with~\eqref{eq:shock_2} reduces by symmetry
to the scalar equation
\begin{equation}
u_{L} \;=\; (h_{\star} - h_{0})\,
\sqrt{\dfrac{g\,(h_{\star} + h_{0})}{2\,h_{\star}\,h_{0}}},
\qquad h_{\star} \;>\; h_{0},
\label{eq:ds_star_match}
\end{equation}
again solved by Brent's algorithm~\cite{ref-brent1971} on the
bracket $(h_{0},\, 500\,h_{0})$. For the Ritter problem no wet
star state exists; only the 1-rarefaction
curve~\eqref{eq:rarefaction_1} parameterized from $(h_{L}, u_{L})$
down toward the dry state $h \to 0^{+}$ is plotted.

The discrete state
$\bigl\{(h_{j}^{N_{t}},\, u_{j}^{N_{t}}) \;:\;
h_{j}^{N_{t}} > 10^{-3}~\mathrm{m}\bigr\}$ is overlaid as point
markers against each case's wave-curve set, sampled at
$N_{c} = 400$ uniform points per curve. A quantitative measure of
confinement to the analytical curves is the Euclidean
nearest-neighbor distance
\begin{equation}
d_{j} \;\equiv\;
\min_{(\tilde{h},\,\tilde{u}) \,\in\, \mathcal{C}}
\sqrt{\bigl(h_{j}^{N_{t}} - \tilde{h}\bigr)^{2}
\;+\; \bigl(u_{j}^{N_{t}} - \tilde{u}\bigr)^{2}},
\label{eq:nn_distance}
\end{equation}
where $\mathcal{C}$ is the union of the sampled wave curves for
the case in question. We report the median, mean, $95$-th
percentile, and maximum of $\{d_{j}\}_{j \in \mathcal{W}}$ over
the wet-cell index set $\mathcal{W} = \{j : h_{j}^{N_{t}}
> 10^{-3}~\mathrm{m}\}$. Because the entropy solution of a
Riemann problem traces $\mathcal{C}$ exactly in the phase plane
at every time, the distribution of $d_{j}$ is the strongest
single-scalar diagnostic of overall solution quality: it
integrates depth and velocity errors with a metric whose units
(m, m\,s$^{-1}$) appear on the two coordinate axes, and any
deficiency of the numerical wave structure, namely excess
dissipation at shocks, incorrect rarefaction-fan slope, or
incomplete relaxation toward the star state, manifests as
displacement off the curves $\mathcal{C}$.

\section{Results and Discussion}

All numerical experiments were executed on a Lenovo ThinkPad T440s
(model 20AQ006HUS) equipped with an Intel Core i7-4600U processor
providing four logical cores at a maximum clock frequency of $3.300$~GHz,
running Linux Lite~6.6 (x86\_64). The four canonical Riemann
configurations were integrated sequentially on a single thread,
completing the full benchmark suite within a four-minute wall-clock
window with zero error and zero warning entries across all four solver
log files. Each simulation was conducted under a common solver
configuration with target CFL number $\mathrm{CFL} = 0.9$,
gravitational acceleration $g = 9.81$~m\,s$^{-2}$, and domain length
$L = 2000$~m, with case-specific cell counts, final times, and
initial conditions as described in the preceding section.
Table~\ref{tab:runtime} summarizes the runtime statistics extracted from
the four solver log files. The adaptive time-step selector saturates
the target CFL number at exactly $0.9000$ in every step of every
run, which is the expected behavior of explicit hyperbolic schemes
whose stability constraint is binding at every
substage~\cite{ref-gottlieb2001}. The solver-reported uncorrected
relative mass change $\Delta M/M(0)$ is exactly zero for the
symmetric quiescent-initial-data Stoker and Ritter cases, in which
no net mass crosses the transmissive boundaries within the
simulated horizon, and is $-24.000\%$ for the diverging
double-rarefaction case and $+24.000\%$ for the converging
double-shock case, reflecting physical mass efflux and influx
through the open boundaries, respectively. The fact that all four
runs completed within one minute of solver-kernel time on a
standard commodity workstation demonstrates that the
Numba-accelerated implementation achieves throughput suitable for
interactive parameter exploration and pedagogical use without
requiring high-performance computing (HPC) resources.

\begin{table}[H]
\centering
\caption{Solver runtime statistics for the four canonical
configurations: cell count $N$, final time $t_{\mathrm{final}}$,
number of adaptive time steps, attained maximum Courant number,
solver-kernel wall-clock time, and uncorrected relative mass
change $\Delta M/M(0) =
\bigl(M(t_{\mathrm{final}}) - M(0)\bigr)/M(0)$. Domain length
$L = 2000$~m, target $\mathrm{CFL} = 0.9$, and $g = 9.81$~m\,
s$^{-2}$ for all runs.}
\label{tab:runtime}
\begin{tabular}{lrrrrrr}
\toprule
Case & $N$ & $t_{\mathrm{final}}$ (s) & Steps & $\max_{n} \mathrm{CFL}^{n}$ & Kernel (s) & $\Delta M/M(0)$ \\
\midrule
Stoker             & 500  & 80 & 283 & 0.9000 & 3.245 & $+0.000\%$  \\
Ritter             & 500  & 40 & 235 & 0.9000 & 0.383 & $+0.000\%$  \\
Double rarefaction & 1000 & 80 & 445 & 0.9000 & 0.598 & $-24.000\%$ \\
Double shock       & 500  & 80 & 188 & 0.9000 & 0.352 & $+24.000\%$ \\
\bottomrule
\end{tabular}
\end{table}

\begin{figure}[H]
\centering
\includegraphics[width=0.95\linewidth]{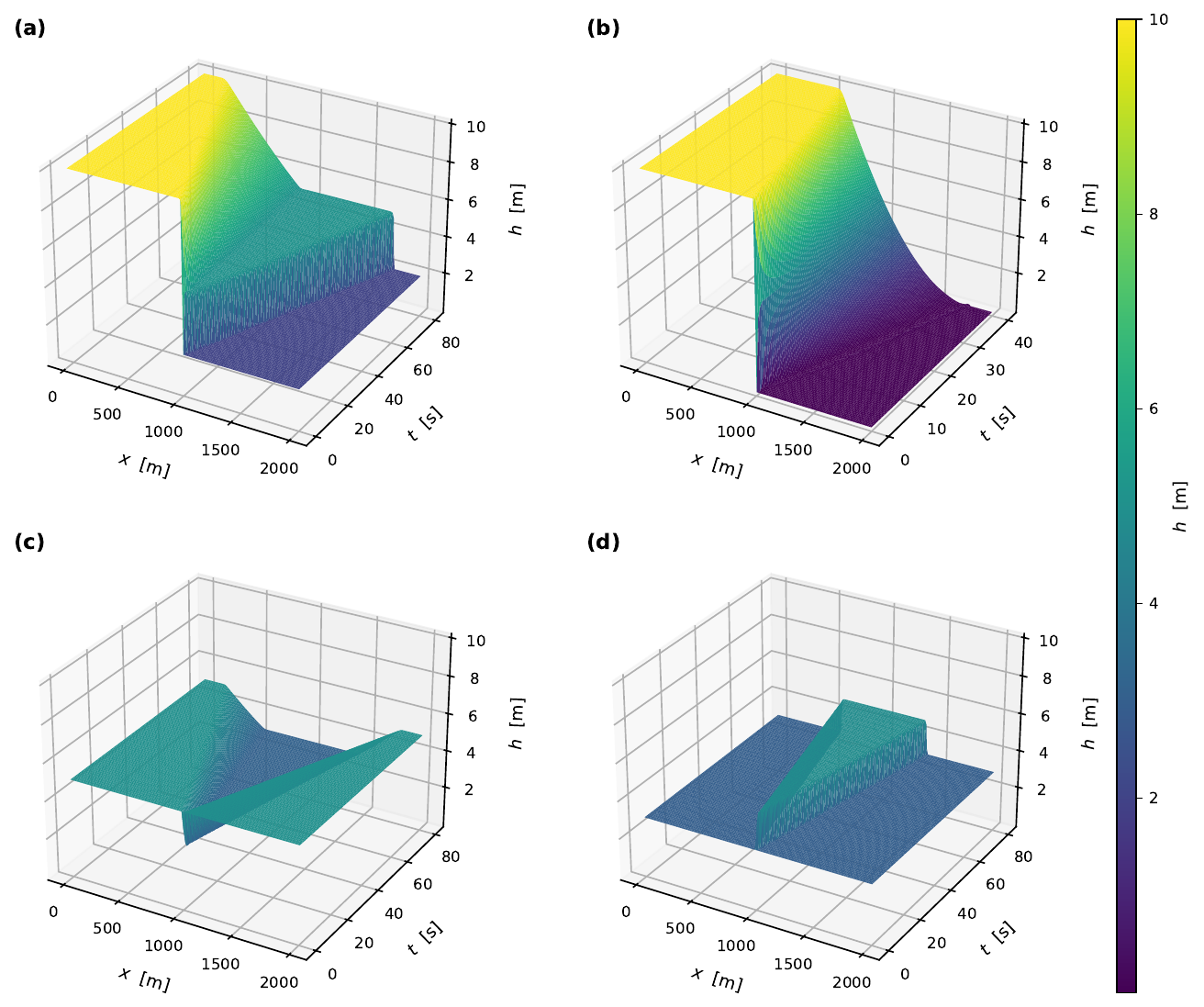}
\caption{Three-dimensional space-time surfaces of the depth field
$h(x, t)$ for the four canonical Riemann configurations:
(a)~Stoker, (b)~Ritter, (c)~double rarefaction, and
(d)~double shock. Common viridis colormap range
$[0.001,\, 10.000]$~m; view elevation $28^{\circ}$, azimuth
$-60^{\circ}$; subsampling $N_{x}^{\mathrm{plot}} = 200$,
$N_{t}^{\mathrm{plot}} = 80$.}
\label{fig:spacetime}
\end{figure}

Figure~\ref{fig:spacetime} renders the spatiotemporal depth field
$h_{j}^{n}$ produced by the finite-volume
update~\eqref{eq:fv_update} as a three-dimensional surface over
$(x, t, h)$ for each of the four cases. The four panels share a
common viridis colormap whose range
$[\min_{c, j, n} h_{j}^{n, c},\,
\max_{c, j, n} h_{j}^{n, c}] = [0.001,\, 10.000]$~m is computed
globally over all four cases $c \in \{1, 2, 3, 4\}$, and a fixed
viewing geometry of elevation $28^{\circ}$ and azimuth
$-60^{\circ}$ is used in every panel. The field is subsampled on a
uniform-index grid of $N_{x}^{\mathrm{plot}} = 200$ spatial nodes
and $N_{t}^{\mathrm{plot}} = 80$ temporal snapshots prior to
rendering.

The qualitative depth-field topology rendered in
Figure~\ref{fig:spacetime} reproduces the canonical elementary
wave structures of the one-dimensional shallow-water Riemann fan
described in~\cite{ref-stoker1957,ref-toro2001}: a left-going
rarefaction connected to a right-going shock through a constant
intermediate plateau for Stoker, a single left-going rarefaction
terminating at a moving dry front for Ritter, two
outward-propagating rarefactions enclosing a central drawdown
for double rarefaction, and two inward-propagating shocks
enclosing a central elevated state for double shock.
Table~\ref{tab:spacetime} reports the quantitative depth-field
statistics extracted from the spatiotemporal arrays for each
case. The extreme values $h_{\min}$ and $h_{\max}$ for the
Stoker and Ritter configurations are attained at the initial
datum $t = 0$~s, namely the $h_{L} = 10$~m water column maximum
at the leftmost cell $x = 2$~m and the immediate-post-dam
minimum at $x = 1002$~m, whereas the double-rarefaction and
double-shock configurations attain their extrema in the interior
of the flow evolution as the wave structure develops: the
central depression to $h_{\min} = 3.011$~m at $t = 1.080$~s in
Figure~\ref{fig:spacetime}c is the deepest drawdown reached by
the diverging rarefactions before they propagate outward, and
the central elevation to $h_{\max} = 4.890$~m at $t = 2.137$~s
in Figure~\ref{fig:spacetime}d is the peak attained at the
moment the two converging shocks first interact. The
spatiotemporal means
$\langle h \rangle_{x, t}$ reflect the integrated effect of the
wave evolution over the full simulation horizon.

\begin{table}[H]
\centering
\caption{Depth-field statistics extracted from the
spatiotemporal arrays rendered in Figure~\ref{fig:spacetime}.
$N_{t}$: number of stored snapshots. $h_{\min}$ and $h_{\max}$:
extreme depth values with their attainment locations
$(x_{\arg\min}, t_{\arg\min})$ and
$(x_{\arg\max}, t_{\arg\max})$.
$\langle h \rangle_{x, t}$: spatiotemporal mean depth.}
\label{tab:spacetime}
\begin{tabular}{lrlll}
\toprule
Case & $N_{t}$ & $h_{\min}$ (m) at $(x\,\mathrm{m},\, t\,\mathrm{s})$ & $h_{\max}$ (m) at $(x\,\mathrm{m},\, t\,\mathrm{s})$ & $\langle h \rangle_{x, t}$ (m) \\
\midrule
Stoker             & 284 & $2.000$ at $(1002,\, 0)$    & $10.000$ at $(2,\, 0)$       & $6.000$  \\
Ritter             & 236 & $0.001$ at $(1002,\, 0)$    & $10.000$ at $(2,\, 0)$       & $5.0005$ \\
Double rarefaction & 446 & $3.011$ at $(999,\, 1.080)$ & $5.000$ at $(1,\, 0)$        & $4.399$  \\
Double shock       & 189 & $3.000$ at $(2,\, 0)$       & $4.890$ at $(998,\, 2.137)$  & $3.361$  \\
\bottomrule
\end{tabular}
\end{table}

\begin{figure}[H]
\centering
\includegraphics[width=0.95\linewidth]{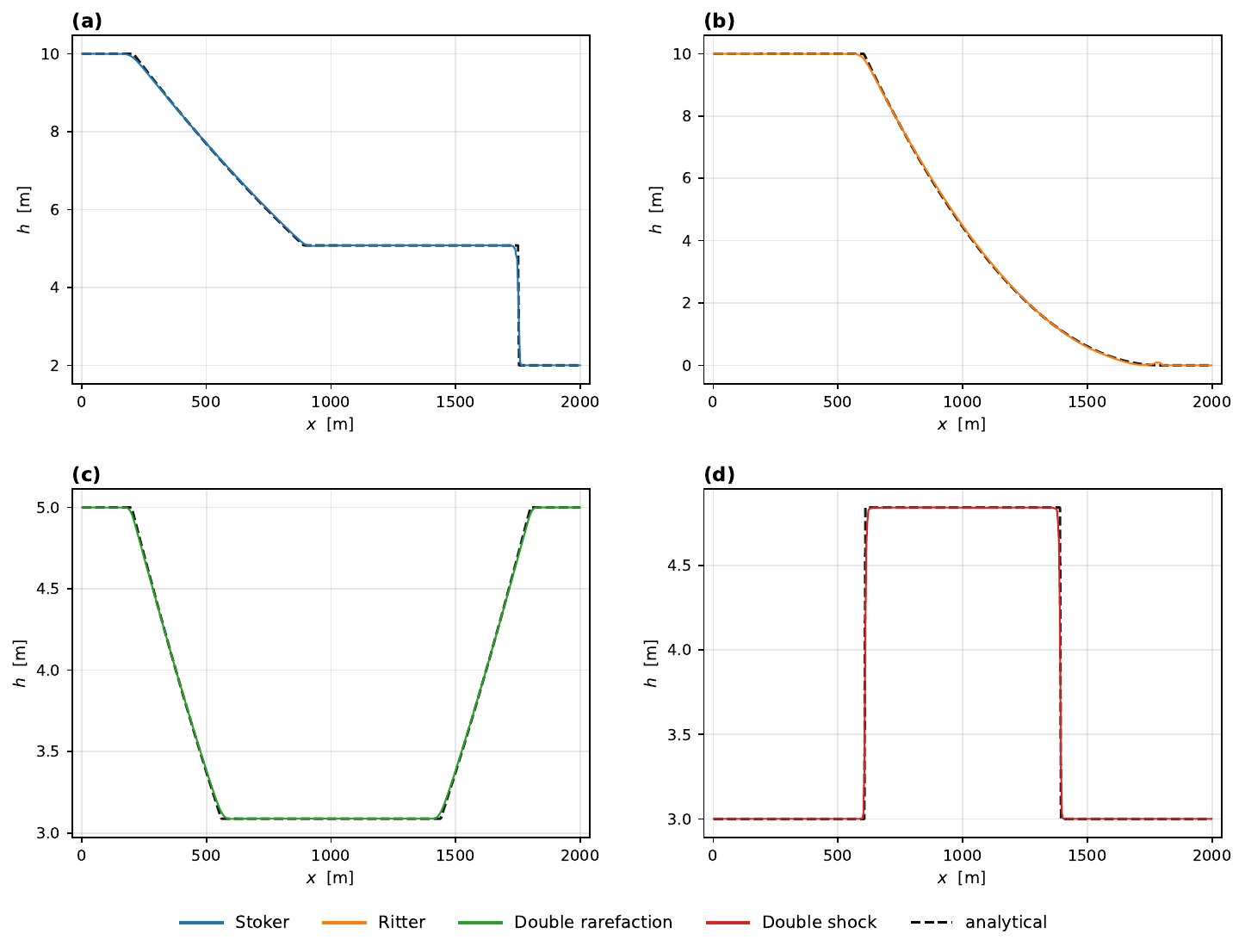}
\caption{Final-time depth profile at $t = t_{\mathrm{final}}$:
numerical solution $h_{\mathrm{num}}(x, t_{\mathrm{final}})$
(colored solid lines) and analytical Riemann solution
$h_{\mathrm{an}}(x, t_{\mathrm{final}})$ (black dashed lines), for
(a)~Stoker ($t_{\mathrm{final}} = 80$~s),
(b)~Ritter ($t_{\mathrm{final}} = 40$~s),
(c)~double rarefaction ($t_{\mathrm{final}} = 80$~s), and
(d)~double shock ($t_{\mathrm{final}} = 80$~s).}
\label{fig:final_time}
\end{figure}

Figure~\ref{fig:final_time} overlays the numerical depth profile
$h_{\mathrm{num}}(x, t_{\mathrm{final}})$ (colored solid line)
and the analytical Riemann solution
$h_{\mathrm{an}}(x, t_{\mathrm{final}})$ (black dashed line) for
each case at $t = t_{\mathrm{final}}$. The pointwise depth error
$e_{j}(t_{\mathrm{final}})$ of~\eqref{eq:pointwise_error} is
summarized through its empirical quantiles
$Q_{p}\bigl[|e_{j}(t_{\mathrm{final}})|\bigr]$ at
$p \in \{50,\, 75,\, 90,\, 95,\, 99,\, 99.9\}$, evaluated by
\texttt{numpy.percentile} with the linear-interpolation
estimator.

Table~\ref{tab:final_time} reports the integral norms, pointwise
error extrema, and empirical error quantiles at
$t = t_{\mathrm{final}}$ for each case. The qualitative pattern
visible in Figure~\ref{fig:final_time} is reflected
quantitatively in the table: the four bulk quantiles
($Q_{50}$ through $Q_{95}$) for Stoker, double rarefaction, and
double shock are smaller than $0.04$~m in every case, while the
two tail quantiles ($Q_{99}$ and $Q_{99.9}$) for Stoker and
double shock are an order of magnitude larger, reaching $1.017$
and $0.6982$~m respectively. This spread between bulk and tail
is characteristic of high-resolution shock-capturing schemes, in
which the formal second-order accuracy of the smooth-region
reconstruction~\eqref{eq:muscl} degrades to first order in the
small set of cells adjacent to a discontinuity by the action of
the slope limiter~\eqref{eq:slope_minmod}~\cite{ref-harten1983,
ref-sweby1984}. For Stoker (Figure~\ref{fig:final_time}a), the
tail quantile of $1.017$~m is localized to the shock-front cells
near $x = 1750$~m where the maximum pointwise error $1.112$~m is
attained, while the median quantile of $5.369 \times 10^{-4}$~m
characterizes the second-order accuracy retained in smooth
regions: the two values span more than three decades. For double
rarefaction (Figure~\ref{fig:final_time}c), whose entropy
solution is everywhere smooth, this separation is largely absent
and the quantile distribution rises monotonically with $p$ from
$3.753 \times 10^{-4}$~m at $p = 50$ to $0.03916$~m at
$p = 99.9$, a span of two decades; the relative pointwise
maximum $\max_{j}|e_{j}|/\langle h \rangle_{x, t}$ in this case
is below $1\%$. The Ritter configuration
(Figure~\ref{fig:final_time}b) is intermediate, with the bulk
of the error distribution concentrated in the rarefaction-fan
cells but no shock-driven tail; the maximum error of $0.156$~m
at $x = 606$~m sits within the developing fan. Use of the
$L^{1}$ norm as the primary integral metric follows from the
fact that the $L^{1}$ topology is the natural metric for entropy
solutions of nonlinear hyperbolic conservation laws, under which
weak solutions exist, are unique, and depend continuously on
initial data~\cite{ref-kruzhkov1970,ref-crandallmajda1980};
the $L^{2}$ values are reported in the table for completeness.

\begin{table}[H]
\centering
\caption{Integral norms, pointwise error extrema, and empirical
error quantiles at $t = t_{\mathrm{final}}$ from
Figure~\ref{fig:final_time}. The quantiles
$Q_{p}\bigl[|e|\bigr]$ are computed by the linear-interpolation
estimator.}
\label{tab:final_time}
\begin{tabular}{lrrrr}
\toprule
Quantity & Stoker & Ritter & Double rar. & Double shock \\
\midrule
$L^{1}(h)(t_{\mathrm{final}})$ (m)            & $28.217$  & $33.091$  & $8.699$               & $11.328$              \\
$L^{2}(h)(t_{\mathrm{final}})$ (m)            & $3.109$   & $1.208$   & $0.389$               & $2.257$               \\
$\max_{j}\,|e_{j}|$ (m)                        & $1.112$   & $0.1564$  & $0.03916$             & $0.6982$              \\
$x_{\arg\max\,|e|}$ (m)                        & $1750$    & $606$     & $1441$                & $610$                 \\
$\langle |e_{j}| \rangle_{j}$ (m)              & $0.01411$ & $0.01655$ & $4.350\times 10^{-3}$ & $5.664\times 10^{-3}$ \\
$Q_{50}\,[\,|e|\,]$ (m)                        & $5.369\times 10^{-4}$ & $0.01095$ & $3.753\times 10^{-4}$ & $0$                   \\
$Q_{75}\,[\,|e|\,]$ (m)                        & $0.01556$             & $0.02822$ & $5.710\times 10^{-3}$ & $4.139\times 10^{-4}$ \\
$Q_{90}\,[\,|e|\,]$ (m)                        & $0.02598$             & $0.03714$ & $0.01460$             & $7.181\times 10^{-4}$ \\
$Q_{95}\,[\,|e|\,]$ (m)                        & $0.03762$             & $0.04253$ & $0.02172$             & $9.510\times 10^{-4}$ \\
$Q_{99}\,[\,|e|\,]$ (m)                        & $0.09536$             & $0.09805$ & $0.03335$             & $0.2178$              \\
$Q_{99.9}\,[\,|e|\,]$ (m)                      & $1.017$               & $0.1553$  & $0.03916$             & $0.6982$              \\
\bottomrule
\end{tabular}
\end{table}

\begin{figure}[H]
\centering
\includegraphics[width=0.95\linewidth]{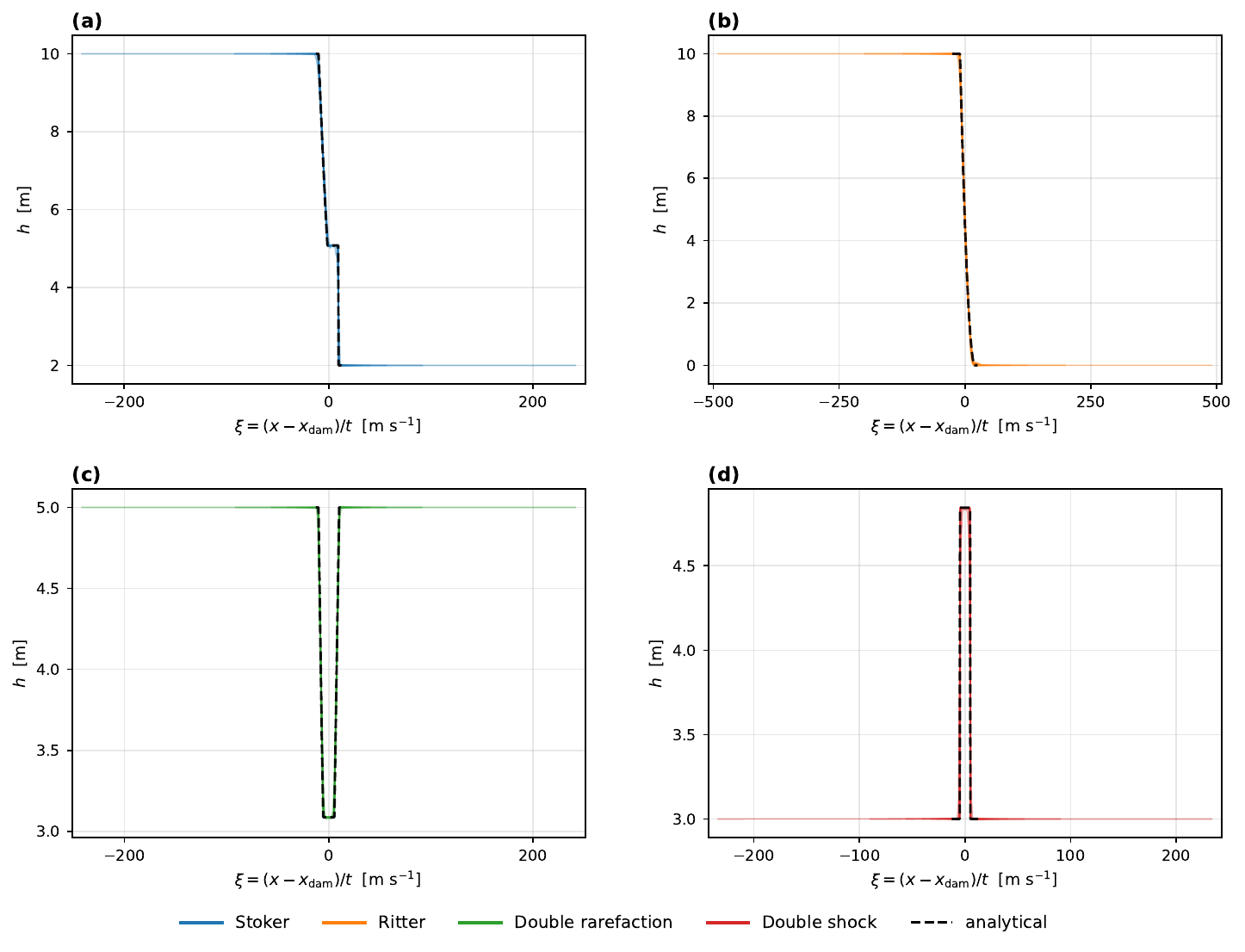}
\caption{Self-similarity collapse of the depth field in the
$(\xi, h)$ plane with similarity variable
$\xi = (x - x_{\mathrm{dam}})/t$. Each panel overlays
$N_{\mathrm{ov}} = 12$ snapshots taken at evenly spaced indices
in the window $t > 0.05\,t_{\mathrm{final}}$ (colored), together
with the analytical curve $H_{\mathrm{an}}(\xi)$ extracted from
the last stored snapshot (black dashed). Panels: (a)~Stoker,
(b)~Ritter, (c)~double rarefaction, (d)~double shock.}
\label{fig:similarity}
\end{figure}

Figure~\ref{fig:similarity} reparameterizes the numerical depth
field as $H_{\mathrm{num}}(\xi)$ via the similarity
variable~\eqref{eq:similarity_var} and overlays
$N_{\mathrm{ov}} = 12$ snapshots per panel taken at evenly
spaced indices within the window
$t > 0.05\,t_{\mathrm{final}}$, together with the analytical
curve $H_{\mathrm{an}}(\xi)$ extracted from the last stored
snapshot. The temporal scatter is summarized by partitioning the
range of $\xi$ values into $N_{b} = 200$ uniform bins and
computing, in each bin containing at least two samples, the
bin-wise sample standard deviation
$\sigma_{b}$~\eqref{eq:sigma_bin}; bins with fewer than two
samples are excluded from the average.

The bin-averaged temporal scatter and the dimensionless
similarity quality~\eqref{eq:Qsim} are reported for each case
in Table~\ref{tab:similarity}. All four configurations attain
$\mathcal{Q}_{\mathrm{sim}}$ above $0.997$, indicating that the
discrete solution preserves the scale invariance of the
continuum Riemann problem to within approximately three parts
in $10^{3}$ over the late-time window. Since the Riemann
problem is invariant under $(x, t) \mapsto (\alpha x,
\alpha t)$ for any $\alpha > 0$, the entropy weak solution
depends on $(x, t)$ only through $\xi$~\cite{ref-lax1957}; the
small residual scatter $\langle \sigma_{b} \rangle_{b}$
therefore quantifies discretization-induced departure from the
exact similarity manifold rather than a structural failure of
the scheme to reproduce it. The double-rarefaction and
double-shock configurations exhibit roughly half the temporal
scatter of Stoker and Ritter, attributable to the fact that
their wave structures do not interact with the dry-bed
degeneracy of~\eqref{eq:sv_eigenvalues} and consequently the
positivity floor~\eqref{eq:positivity_fix} never activates over
the simulation horizon.

\begin{table}[H]
\centering
\caption{Self-similarity collapse statistics from
Figure~\ref{fig:similarity}: bin-averaged temporal scatter
$\langle \sigma_{b} \rangle_{b}$, spatiotemporal mean depth
$\langle h \rangle$, and dimensionless similarity quality
$\mathcal{Q}_{\mathrm{sim}}$.}
\label{tab:similarity}
\begin{tabular}{lrrr}
\toprule
Case & $\langle \sigma_{b} \rangle_{b}$ (m) & $\langle h \rangle$ (m) & $\mathcal{Q}_{\mathrm{sim}}$ \\
\midrule
Stoker             & $0.01169$              & $6.000$  & $0.99805$ \\
Ritter             & $0.01471$              & $5.0005$ & $0.99706$ \\
Double rarefaction & $5.682 \times 10^{-3}$ & $4.399$  & $0.99871$ \\
Double shock       & $5.024 \times 10^{-3}$ & $3.361$  & $0.99850$ \\
\bottomrule
\end{tabular}
\end{table}

\begin{figure}[H]
\centering
\includegraphics[width=0.95\linewidth]{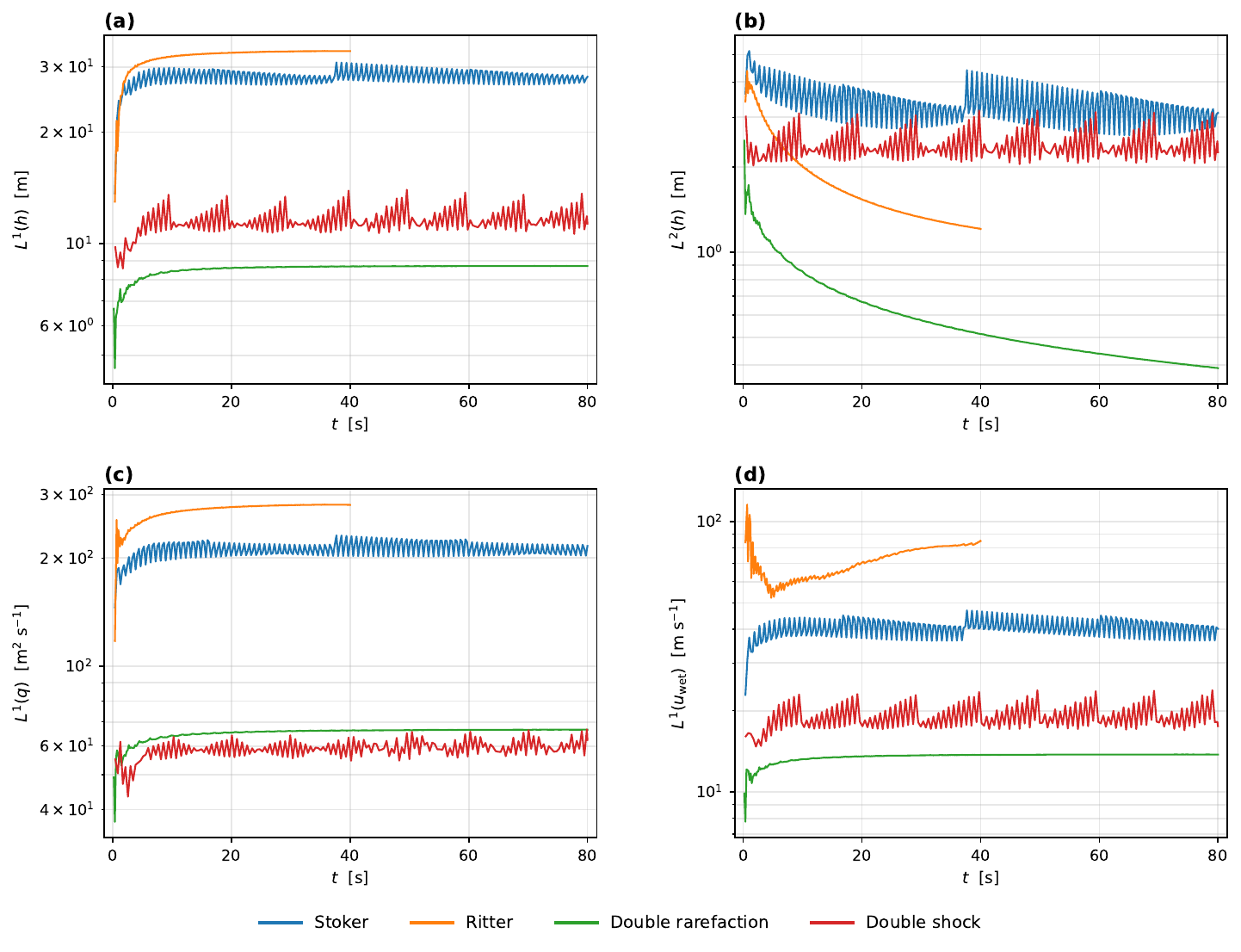}
\caption{Time evolution of integral error norms on a
logarithmic ordinate against the exact Riemann solution:
(a)~$L^{1}(h)$, (b)~$L^{2}(h)$, (c)~$L^{1}(q)$,
(d)~$L^{1}(u_{\mathrm{wet}})$, with the wet-cell restriction in
(d) applied to cells satisfying
$h_{j}^{n} > H_{\mathrm{DRY}} = 10^{-2}$~m on both the
numerical and analytical sides. Curves: Stoker (blue), Ritter
(orange), double rarefaction (green), double shock (red).}
\label{fig:norms}
\end{figure}

Figure~\ref{fig:norms} plots the four integral error
norms~\eqref{eq:Lp_norm} computed by the solver against the
exact Riemann solution at every stored time step: $L^{1}(h)$ in
Figure~\ref{fig:norms}a, $L^{2}(h)$ in Figure~\ref{fig:norms}b,
$L^{1}(q)$ in Figure~\ref{fig:norms}c, and
$L^{1}(u_{\mathrm{wet}})$ in Figure~\ref{fig:norms}d, the last
restricted to the wet-cell index set~\eqref{eq:wet_set} with
$H_{\mathrm{DRY}} = 10^{-2}$~m. All four cases are overlaid on
every panel and identified by color, and the ordinate is
logarithmic in all four panels. The total number of stored
snapshots is $284$ for Stoker, $236$ for Ritter, $446$ for
double rarefaction, and $189$ for double shock.

Table~\ref{tab:norms} reports the final-time value, envelope
maximum over $t > 0$, and time mean over $t > 0$ for each of
the four error norms. The $L^{1}(h)$ time series in
Figure~\ref{fig:norms}a saturate to a near-constant level after
the initial transient, with the time mean within $4\%$ of the
envelope maximum for Stoker, Ritter, and double rarefaction;
this saturation is characteristic of second-order
total-variation-diminishing schemes applied to Riemann problems
containing discontinuities, where the dominant error
contribution scales as $\mathcal{O}(\Delta x)$ at each shock or
rarefaction edge and persists at fixed cell
count~\cite{ref-harten1983,ref-sweby1984,ref-leveque2002}. The
$L^{2}(h)$ time series for Ritter and double rarefaction in
Figure~\ref{fig:norms}b exhibit early-time decay from a
transient peak ($4.380$~m and $2.481$~m respectively) to a
stationary value ($1.208$~m and $0.389$~m), reflecting
smoothing of the initial discontinuity into the self-similar
fan once the entropy weak solution has been established. The
choice of $L^{1}(q)$ rather than $L^{1}(u)$ as the primary
momentum-error diagnostic is motivated by the divergence of
$u = q/h$ as the positivity floor~\eqref{eq:positivity_fix} is
approached: for Ritter, the solver-stored raw velocity error
$L^{1}(u)$ reaches $322.76$~m\,s$^{-1}$ at
$t_{\mathrm{final}}$, whereas the wet-cell restriction yields
$L^{1}(u_{\mathrm{wet}}) = 84.753$~m\,s$^{-1}$, and the
discharge norm $L^{1}(q) = 281.080$~m$^{2}$\,s$^{-1}$ provides
a representation that remains bounded throughout the dry-front
evolution. Analogous issues motivate the well-balanced and
augmented Riemann-solver frameworks developed for SWE solvers
admitting bed-slope and wetting-and-drying source
terms~\cite{ref-audusse2004,ref-george2008}.

\begin{table}[H]
\centering
\caption{Integral error norms from Figure~\ref{fig:norms}:
final-time value at $t = t_{\mathrm{final}}$, envelope maximum
over $t > 0$, and time mean over $t > 0$. ``DR'' denotes double
rarefaction, ``DS'' denotes double shock.}
\label{tab:norms}
\begin{tabular}{llrrrr}
\toprule
Norm & Statistic & Stoker & Ritter & DR & DS \\
\midrule
$L^{1}(h)$ (m)
  & at $t_{\mathrm{final}}$         & $28.217$  & $33.091$  & $8.699$  & $11.328$ \\
  & $\max_{t}$                       & $30.858$  & $33.130$  & $8.709$  & $13.985$ \\
  & $\langle \cdot \rangle_{t}$      & $28.144$  & $31.876$  & $8.559$  & $11.508$ \\
\midrule
$L^{2}(h)$ (m)
  & at $t_{\mathrm{final}}$         & $3.109$   & $1.208$   & $0.389$  & $2.257$  \\
  & $\max_{t}$                       & $5.150$   & $4.380$   & $2.481$  & $3.172$  \\
  & $\langle \cdot \rangle_{t}$      & $3.368$   & $1.830$   & $0.599$  & $2.393$  \\
\midrule
$L^{1}(q)$ (m$^{2}$\,s$^{-1}$)
  & at $t_{\mathrm{final}}$         & $216.014$ & $281.080$ & $66.664$ & $62.410$ \\
  & $\max_{t}$                       & $231.225$ & $281.640$ & $66.726$ & $66.809$ \\
  & $\langle \cdot \rangle_{t}$      & $211.393$ & $269.657$ & $65.420$ & $58.996$ \\
\midrule
$L^{1}(u_{\mathrm{wet}})$ (m\,s$^{-1}$)
  & at $t_{\mathrm{final}}$         & $40.164$  & $84.753$  & $13.773$ & $17.537$ \\
  & $\max_{t}$                       & $46.988$  & $115.600$ & $13.788$ & $23.814$ \\
  & $\langle \cdot \rangle_{t}$      & $40.458$  & $71.046$  & $13.529$ & $18.984$ \\
\bottomrule
\end{tabular}
\end{table}

\begin{figure}[H]
\centering
\includegraphics[width=0.95\linewidth]{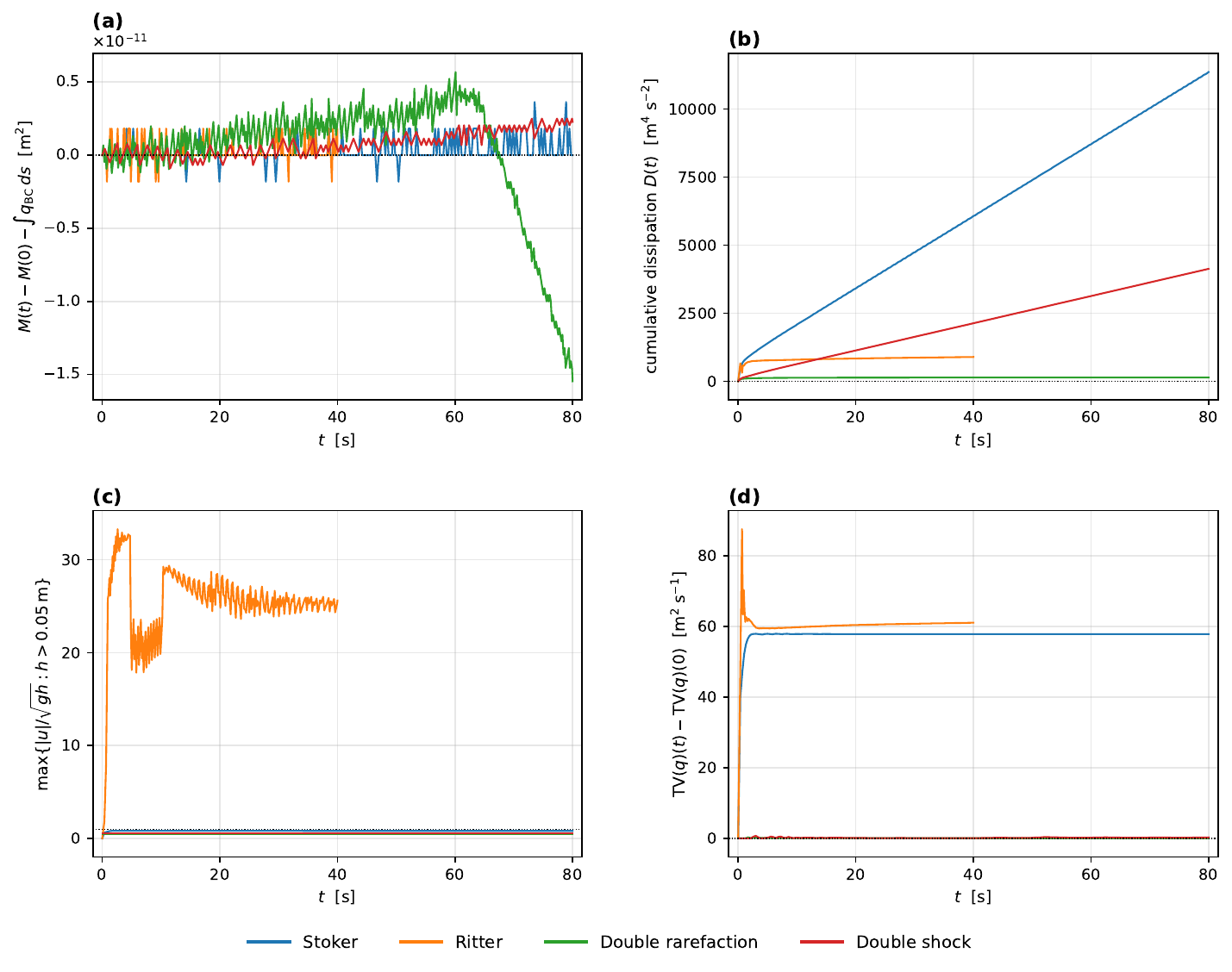}
\caption{Boundary-flux-corrected conservation diagnostics.
(a)~Mass residual
$r_{M}(t) = M(t) - M(0) - \Phi_{M}(t)$ on a scale of
$10^{-11}$~m$^{2}$;
(b)~cumulative dissipation
$D(t) = \Phi_{E}(t) - (E(t) - E(0))$;
(c)~strict-mask wet-cell maximum Froude number
$\max\{|u|/\!\sqrt{g\,h} : h > H_{\mathrm{strict}}\}$ with
$H_{\mathrm{strict}} = 0.05$~m, dotted reference at unity;
(d)~total-variation growth
$\mathrm{TV}(q)(t) - \mathrm{TV}(q)(0)$. Curves: Stoker (blue),
Ritter (orange), double rarefaction (green), double shock
(red).}
\label{fig:conservation}
\end{figure}

Figure~\ref{fig:conservation} reports four
boundary-flux-corrected conservation diagnostics with strict
wet-cell positivity threshold
$H_{\mathrm{strict}} = 5\,H_{\mathrm{DRY}} = 0.05$~m.
Figure~\ref{fig:conservation}a plots the mass residual
$r_{M}(t)$~\eqref{eq:mass_residual} on a scale of
$10^{-11}$~m$^{2}$ with a dotted reference at zero;
Figure~\ref{fig:conservation}b plots the cumulative dissipation
$D(t)$~\eqref{eq:dissipation} with a dotted reference at zero;
Figure~\ref{fig:conservation}c plots the strict-mask wet-cell
maximum Froude number~\eqref{eq:strict_froude} with a dotted
reference at unity; and Figure~\ref{fig:conservation}d plots
the total-variation growth
$\Delta\mathrm{TV}(q)(t)$~\eqref{eq:dTVq} of the conserved
discharge above its initial value, with a dotted reference at
zero. All four cases are overlaid on every panel and identified
by color.

Table~\ref{tab:conservation_mass} reports the mass-closure and
energy-budget diagnostics, and Table~\ref{tab:conservation_frtv}
reports the Froude and total-variation diagnostics, for each
case. The normalized mass residual $\max_{n}|r_{M}^{n}|/M(0)$
is below $1.6 \times 10^{-15}$ in every case, which is at the
level of accumulated floating-point round-off for
double-precision arithmetic; this confirms that the
finite-volume update~\eqref{eq:fv_update} is exactly
mass-conserving up to machine precision once the boundary-flux
contribution~\eqref{eq:mass_inflow} is subtracted, as required
by the Lax--Wendroff structure of conservative
discretizations~\cite{ref-leveque2002}. The cumulative
dissipation $D(t^{n})$ is non-negative for every $n$ in every
case, with $\min_{n} D(t^{n}) = 0$ identically, satisfying the
discrete entropy-admissibility
condition~\eqref{eq:entropy_admissibility} expected of any
consistent shock-capturing scheme on the SWE conservation
law~\cite{ref-lax1957,ref-kruzhkov1970,ref-tadmor1987}. The
relative dissipation $D(t_{\mathrm{final}})/E(0)$ varies from
$0.047\%$ for double rarefaction (smooth solution, near-floor
numerical dissipation) and $0.182\%$ for Ritter (a
rarefaction-only entropy solution, with the small remaining
dissipation attributable to numerical viscosity at the
dry-front interface), through $2.227\%$ for Stoker (single
right-going shock), up to $3.582\%$ for double shock (two
converging shocks); this monotone ordering by wave-type
content is consistent with shocks providing the dominant
mechanism of mechanical-energy loss. The strict-mask wet-cell
Froude peaks reported in Table~\ref{tab:conservation_frtv}
place Stoker, double rarefaction, and double shock entirely in
the subcritical regime, while the Ritter peak of $33.287$ with
$99.58\%$ supercritical fraction reflects the transcritical
character of the dry-bed Riemann fan, in which $u(\xi)$ remains
$\mathcal{O}(c_{L})$ but $h(\xi) \to 0^{+}$ at the leading edge
of the rarefaction; this regime is precisely the one addressed
by augmented Riemann solvers designed for SWE problems with
wetting-and-drying interfaces~\cite{ref-george2008}. The
total-variation growth above the initial value is
$0.301$~m$^{2}$\,s$^{-1}$ for double rarefaction (which is
$1.00\%$ of $\mathrm{TV}(q)(0) = 30.000$~m$^{2}$\,s$^{-1}$) and
$0.729$~m$^{2}$\,s$^{-1}$ for double shock (which is $4.05\%$
of $\mathrm{TV}(q)(0) = 18.000$~m$^{2}$\,s$^{-1}$), consistent
with the empirical total-variation-diminishing property of
MUSCL--minmod reconstruction combined with SSP--RK2 time
stepping for the nonlinear SWE
system~\cite{ref-harten1983,ref-sweby1984,ref-gottlieb2001}.
For Stoker and Ritter, $\mathrm{TV}(q)(0) = 0$ because the
initial discharge field $q(x, 0) \equiv 0$ is uniform; the
growth to $57.851$~m$^{2}$\,s$^{-1}$ and
$61.089$~m$^{2}$\,s$^{-1}$ at $t = t_{\mathrm{final}}$
reflects the development of the analytical Riemann fan, whose
own total variation is finite and positive at every $t > 0$,
and is therefore physically required rather than a violation
of the TVD bound.

\begin{table}[H]
\centering
\caption{Mass-closure and energy-budget diagnostics from
Figure~\ref{fig:conservation} (panels a, b): discrete
invariants at $t = 0$ and $t = t_{\mathrm{final}}$,
cumulative boundary flux, maximum absolute BC-corrected
residual, and dissipation. ``DR'' denotes double rarefaction,
``DS'' denotes double shock.}
\label{tab:conservation_mass}
\begin{tabular}{llrrrr}
\toprule
Block & Quantity & Stoker & Ritter & DR & DS \\
\midrule
Mass
  & $M(0)$ (m$^{2}$)                               & $1.2000 \times 10^{4}$  & $1.0001 \times 10^{4}$  & $1.0000 \times 10^{4}$    & $6.000 \times 10^{3}$  \\
  & $M(t_{\mathrm{final}})$ (m$^{2}$)              & $1.2000 \times 10^{4}$  & $1.0001 \times 10^{4}$  & $7.600 \times 10^{3}$     & $7.440 \times 10^{3}$  \\
  & $\Phi_{M}(t_{\mathrm{final}})$ (m$^{2}$)       & $1.504 \times 10^{-15}$ & $0$                     & $-2.400 \times 10^{3}$    & $1.440 \times 10^{3}$  \\
  & $\max_{n}|r_{M}^{n}|$ (m$^{2}$)                & $3.638 \times 10^{-12}$ & $1.819 \times 10^{-12}$ & $1.546 \times 10^{-11}$   & $2.501 \times 10^{-12}$ \\
  & $\max_{n}|r_{M}^{n}|/M(0)$                     & $3.032 \times 10^{-16}$ & $1.819 \times 10^{-16}$ & $1.546 \times 10^{-15}$   & $4.169 \times 10^{-16}$ \\
\midrule
Energy
  & $E(0)$ (m$^{4}$\,s$^{-2}$)                     & $5.1012 \times 10^{5}$  & $4.9050 \times 10^{5}$  & $2.9025 \times 10^{5}$    & $1.1529 \times 10^{5}$ \\
  & $E(t_{\mathrm{final}})$ (m$^{4}$\,s$^{-2}$)    & $4.9876 \times 10^{5}$  & $4.8961 \times 10^{5}$  & $1.6159 \times 10^{5}$    & $1.6002 \times 10^{5}$ \\
  & $\Phi_{E}(t_{\mathrm{final}})$ (m$^{4}$\,s$^{-2}$) & $1.475 \times 10^{-13}$ & $0$                & $-1.2852 \times 10^{5}$   & $4.8859 \times 10^{4}$ \\
  & $D(t_{\mathrm{final}})$ (m$^{4}$\,s$^{-2}$)    & $1.1363 \times 10^{4}$  & $891.489$               & $137.083$                 & $4.1301 \times 10^{3}$ \\
  & $\min_{n} D(t^{n})$ (m$^{4}$\,s$^{-2}$)        & $0$                     & $0$                     & $0$                       & $0$                    \\
\bottomrule
\end{tabular}
\end{table}

\begin{table}[H]
\centering
\caption{Strict-mask wet-cell maximum Froude number and
total-variation diagnostics from Figure~\ref{fig:conservation}
(panels c, d). The supercritical fraction is the fraction of
stored snapshots for which
$\mathrm{Fr}_{\mathrm{strict}}^{n} > 1$. Strict-mask threshold
$H_{\mathrm{strict}} = 5\,H_{\mathrm{DRY}} = 0.05$~m.}
\label{tab:conservation_frtv}
\begin{tabular}{lrrrr}
\toprule
Quantity & Stoker & Ritter & DR & DS \\
\midrule
$\max_{n} \mathrm{Fr}_{\mathrm{strict}}^{n}$               & $0.828$  & $33.287$ & $0.428$  & $0.553$  \\
Supercritical fraction                                       & $0$      & $0.9958$ & $0$      & $0$      \\
$\mathrm{TV}(q)(0)$ (m$^{2}$\,s$^{-1}$)                     & $0$      & $0$      & $30.000$ & $18.000$ \\
$\mathrm{TV}(q)(t_{\mathrm{final}})$ (m$^{2}$\,s$^{-1}$)    & $57.851$ & $61.089$ & $30.037$ & $18.277$ \\
$\max_{n} \Delta\mathrm{TV}(q)(t^{n})$ (m$^{2}$\,s$^{-1}$)  & $57.967$ & $87.578$ & $0.301$  & $0.729$  \\
\bottomrule
\end{tabular}
\end{table}

\begin{figure}[H]
\centering
\includegraphics[width=0.95\linewidth]{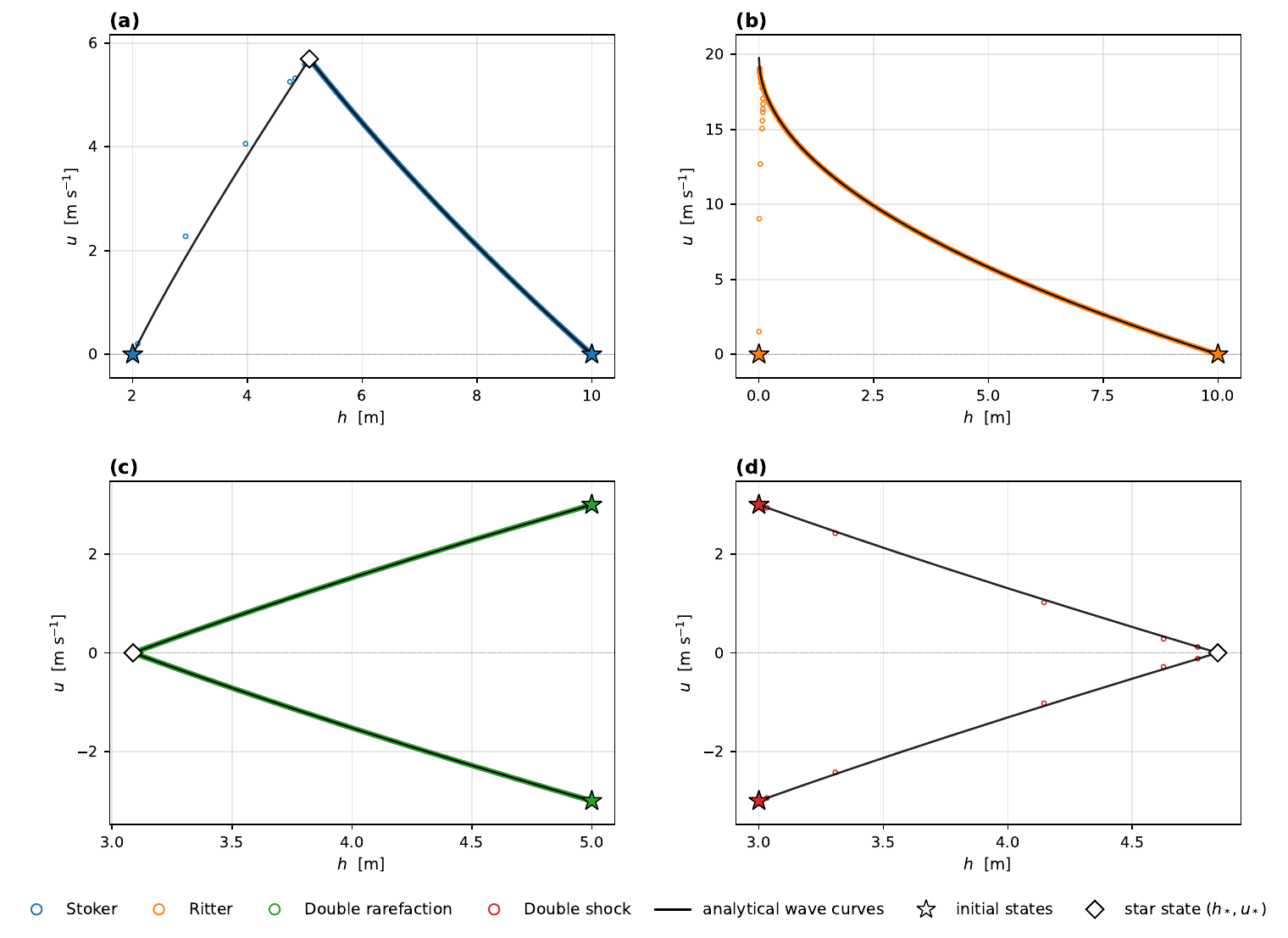}
\caption{Phase-plane projection of the final-time numerical
state $\{(h_{j}^{N_{t}},\, u_{j}^{N_{t}}) :
h_{j}^{N_{t}} > 10^{-3}~\mathrm{m}\}$ (colored open markers)
overlaid on the analytical wave curves (black solid) sampled at
$N_{c} = 400$ points per curve. Filled stars mark the initial
states $(h_{L}, u_{L})$ and $(h_{R}, u_{R})$; open diamonds
mark the analytical star state $(h_{\star}, u_{\star})$ where
one exists. Panels: (a)~Stoker, (b)~Ritter, (c)~double
rarefaction, (d)~double shock.}
\label{fig:phase}
\end{figure}

Figure~\ref{fig:phase} displays the final-time numerical state
$\bigl\{(h_{j}^{N_{t}},\, u_{j}^{N_{t}}) :
h_{j}^{N_{t}} > 10^{-3}~\mathrm{m}\bigr\}$ as colored open
markers in the conserved-variable plane $(h, u)$ for each case,
overlaid against the analytical wave curves of the Riemann
fan~\eqref{eq:rarefaction_1}--\eqref{eq:shock_2} drawn as black
solid lines, with $N_{c} = 400$ uniform points sampled per
wave curve. Filled stars mark the initial states
$(h_{L}, u_{L})$ and $(h_{R}, u_{R})$, and open diamonds mark
the analytical star state $(h_{\star}, u_{\star})$ where one
exists. The Euclidean nearest-neighbor distance
$d_{j}$~\eqref{eq:nn_distance} from each numerical cell to the
nearest sampled point of the wave-curve set is summarized
through its median, mean, $95$-th percentile, and maximum over
the wet-cell index set.

Table~\ref{tab:phase} reports, for each case, the number of
final-time cells exceeding the phase-space filter, the
analytical star-state coordinates where defined, and the
distribution statistics of the Euclidean distance $d_{j}$. The
median Euclidean distance is below $10^{-2}$ in every
configuration, and below $10^{-3}$ in three of the four
configurations, demonstrating that the discrete solution
remains confined to the integral manifold spanned by the
Riemann invariants~\eqref{eq:riemann_invariants} and the
Rankine--Hugoniot shock
curves~\eqref{eq:shock_1}--\eqref{eq:shock_2} to within
sub-centimetre accuracy in $(h, u)$~\cite{ref-lax1957,
ref-toro2001}. The larger maximum distance of $7.476$ for
Ritter (Figure~\ref{fig:phase}b) is concentrated in cells
immediately adjacent to the dry front where the positivity
floor~\eqref{eq:positivity_fix} activates and the analytical
$1$-rarefaction curve approaches the singular limit
$h \to 0^{+}$; these contributions are absent from the median
statistic, which remains at $9.703 \times 10^{-3}$, and are
attributable to the same dry-bed degeneracy that also produces
the supercritical Froude excursion of
Figure~\ref{fig:conservation}c. The numerically determined
star states for Stoker, double rarefaction, and double shock
agree with the analytical values
of~\eqref{eq:stoker_star_match}, \eqref{eq:dr_star}, and
\eqref{eq:ds_star_match} to within the $95$-th percentile
distances of $8.804 \times 10^{-3}$, $3.800 \times 10^{-3}$,
and $1.591 \times 10^{-3}$ respectively. Taken together, the
four configurations exhaust the elementary one-dimensional
wave structures admissible to~\eqref{eq:sv_conservative} in
the absence of source terms and constitute a verification
suite comparable in scope to that compiled in the SWASHES
library of analytical solutions for shallow-water solver
benchmarking~\cite{ref-delestre2013}, providing reference
numerical outputs against which extensions of \texttt{amerta}
incorporating bed slope, friction, or two-dimensional
generalization can be tested.

\begin{table}[H]
\centering
\caption{Phase-plane statistics from Figure~\ref{fig:phase}:
number of final-time cells $\#\{j : h_{j}^{N_{t}} > 10^{-3}\}$
exceeding the phase-space filter, analytical star-state
coordinates $(h_{\star}, u_{\star})$ where defined, and
distribution statistics of the Euclidean distance $\{d_{j}\}$
from each cell to the nearest sampled analytical wave curve.}
\label{tab:phase}
\begin{tabular}{lrrrr}
\toprule
Quantity & Stoker & Ritter & DR & DS \\
\midrule
Active cells / $N$                 & $500/500$              & $456/500$              & $1000/1000$             & $500/500$              \\
$h_{\star}$ (m)                    & $5.0787$               & n.a.                   & $3.0876$                & $4.8437$               \\
$u_{\star}$ (m\,s$^{-1}$)          & $5.6921$               & n.a.                   & $0$                     & $0$                    \\
$\mathrm{median}\,\{d_{j}\}$       & $9.102 \times 10^{-4}$ & $9.703 \times 10^{-3}$ & $4.705 \times 10^{-4}$  & $0$                    \\
$\mathrm{mean}\,\{d_{j}\}$         & $3.366 \times 10^{-3}$ & $0.1096$               & $1.134 \times 10^{-3}$  & $7.578 \times 10^{-4}$ \\
$Q_{95}\,\{d_{j}\}$                & $8.804 \times 10^{-3}$ & $0.1239$               & $3.800 \times 10^{-3}$  & $1.591 \times 10^{-3}$ \\
$\max\,\{d_{j}\}$                  & $0.1875$               & $7.476$                & $4.616 \times 10^{-3}$  & $0.02948$              \\
\bottomrule
\end{tabular}
\end{table}

\section{Conclusions}

This work has introduced \texttt{amerta}, an open-source Python
implementation of a one-dimensional Saint--Venant solver based on a
MUSCL--HLLC reconstruction with SSP--RK2 time integration and Numba
JIT compilation of the performance-critical kernels. The
library is deliberately restricted to the frictionless
prismatic-channel limit of the conservative form so that solver
behavior can be evaluated end-to-end against closed-form analytical
solutions of the canonical wet-bed and dry-bed Riemann problems, and
is distributed under the MIT license with the four benchmark
configurations built in alongside a programmatic interface for
user-defined initial data that is configurable through plain-text
files without modification of the solver source. Future development
of the code, comprising the incorporation of bed slope, Manning or
Ch\'{e}zy friction, channel-geometry variability, lateral mass and
momentum source terms, and two-dimensional generalization on
unstructured or curvilinear meshes, is anticipated to require
minimal modification of the conservative finite-volume kernel; for
each such extension, the source-free reduction must continue to
reproduce the canonical Riemann fan structures examined here, and
the diagnostic suite developed in this work can be reused without
modification to verify the source-free behavior of the extended
solver against the same analytical benchmarks. The complete source
code, configuration files, analytical-solution evaluators,
post-processing scripts, NetCDF archives of the four benchmark
runs, and the figures and statistics files generated for this paper
are openly available through the public software and data
repositories of the project.

\section*{Acknowledgements}
The authors utilized Claude Sonnet~4.6 (Anthropic, PBC) solely as a
writing-assistance tool to refine English vocabulary and grammar during
the preparation of this manuscript. All scientific content,
interpretations, analyses, conclusions, and any remaining linguistic
imperfections are the sole responsibility of the authors.

\section*{Funding}
This study was funded by ITB Research Program under ITB Flagship Research Scheme through the Directorate of Research and Innovation, Bandung Institute of Technology. 

\section*{Author Contributions}
\textbf{D.E.I.}:
conceptualization, methodology, funding acquisition, project
administration, supervision, writing - original draft preparation,
writing - review and editing. \textbf{S.H.S.H.}: conceptualization,
methodology, software, formal analysis, investigation, data
curation, validation, visualization, writing - original draft
preparation. \textbf{I.P.A.}: methodology, supervision, writing -
review and editing. \textbf{F.K.}: methodology, supervision,
writing - review and editing. \textbf{A.P.}: validation, writing -
review and editing. \textbf{R.D.K.}: validation, writing - review
and editing. \textbf{E.R.}: supervision, writing - review and
editing. \textbf{R.S.}: supervision, writing - review and editing.
\textbf{D.J.P.}: funding acquisition, supervision, writing - review
and editing. All authors have read and agreed to the published
version of the manuscript.

\section*{Data Availability}
This work adheres to the Findable, Accessible, Interoperable,
Reusable (FAIR) principles, with all artifacts released under the MIT
license. The \texttt{amerta} solver source code is hosted on GitHub
at \url{https://github.com/sandyherho/amerta} and distributed
through the Python Package Index (PyPI) at
\url{https://pypi.org/project/amerta/}. The post-processing scripts
that generate every figure and statistical diagnostic reported here
are archived at \url{https://github.com/sandyherho/suppl_amerta}.
The complete computational outputs of the four benchmark runs,
comprising log files, NetCDF archives, statistics files, animated
visualizations, and rendered figures, are permanently deposited on
the Open Science Framework (OSF) at
\url{https://doi.org/10.17605/OSF.IO/MT8GK}.


\begin{thebibliography}{999}

\bibitem{ref-svc1871}
de Saint-Venant, A.J.C. Th\'{e}orie du Mouvement Non Permanent des Eaux,
avec Application aux Crues des Rivi\`{e}res et \`{a} l'Introduction des
Mar\'{e}es dans leur Lit. \textit{Comptes Rendus Acad. Sci. Paris}
\textbf{1871}, \textit{73}, 147--154, 237--240.

\bibitem{ref-ritter1892}
Ritter, A. Die Fortpflanzung der Wasserwellen.
\textit{Z. Ver. Deutsch. Ing.} \textbf{1892}, \textit{36}, 947--954.

\bibitem{ref-stoker1957}
Stoker, J.J. \textit{Water Waves: The Mathematical Theory with
Applications}; Interscience Publishers: New York, NY, USA, 1957.

\bibitem{ref-lax1957}
Lax, P.D. Hyperbolic systems of conservation laws II.
\textit{Commun. Pure Appl. Math.} \textbf{1957}, \textit{10},
537--566. \url{https://doi.org/10.1002/cpa.3160100406}.

\bibitem{ref-batchelor1967}
Batchelor, G.K. \textit{An Introduction to Fluid Dynamics};
Cambridge University Press: Cambridge, UK, 1967.

\bibitem{ref-kruzhkov1970}
Kruzhkov, S.N. First Order Quasilinear Equations in Several
Independent Variables. \textit{Math. USSR Sb.} \textbf{1970},
\textit{10}, 217--243.
\url{https://doi.org/10.1070/SM1970v010n02ABEH002156}.

\bibitem{ref-brent1971}
Brent, R.P. An algorithm with guaranteed convergence for finding
a zero of a function. \textit{Comput. J.} \textbf{1971},
\textit{14(4)}, 422--425.
\url{https://doi.org/10.1093/comjnl/14.4.422}.

\bibitem{ref-flanders1973}
Flanders, H. Differentiation Under the Integral Sign.
\textit{Am. Math. Mon.} \textbf{1973}, \textit{80(6)}, 615--627.
\url{https://doi.org/10.2307/2319163}.

\bibitem{ref-vanleer1979}
van Leer, B. Towards the Ultimate conservative difference scheme. V.
A second-Order sequel to Godunov's Method.
\textit{J. Comput. Phys.} \textbf{1979}, \textit{32(1)}, 101--136.
\url{https://doi.org/10.1016/0021-9991(79)90145-1}.

\bibitem{ref-crandallmajda1980}
Crandall, M.G.; Majda, A. Monotone difference approximations for
scalar conservation laws. \textit{Math. Comp.} \textbf{1980},
\textit{34}, 1--21.
\url{https://doi.org/10.1090/S0025-5718-1980-0551288-3}.

\bibitem{ref-hartenlaxleer1983}
Harten, A.; Lax, P.D.; van Leer, B. On Upstream Differencing and
Godunov-Type Schemes for Hyperbolic Conservation Laws.
\textit{SIAM Rev.} \textbf{1983}, \textit{25(1)}, 35--61.
\url{https://doi.org/10.1137/1025002}.

\bibitem{ref-harten1983}
Harten, A. High resolution schemes for hyperbolic conservation
laws. \textit{J. Comput. Phys.} \textbf{1983}, \textit{49(3)},
357--393.
\url{https://doi.org/10.1016/0021-9991(83)90136-5}.

\bibitem{ref-sweby1984}
Sweby, P.K. High Resolution Schemes Using Flux Limiters for
Hyperbolic Conservation Laws. \textit{SIAM J. Numer. Anal.}
\textbf{1984}, \textit{21(5)}, 995--1011.
\url{https://doi.org/10.1137/0721062}.

\bibitem{ref-landau1987}
Landau, L.D.; Lifshitz, E.M. \textit{Fluid Mechanics}, 2nd ed.;
Pergamon Press: Oxford, UK, 1987; Vol.~6 of Course of Theoretical
Physics.

\bibitem{ref-tadmor1987}
Tadmor, E. The numerical viscosity of entropy stable schemes for
systems of conservation laws. I. \textit{Math. Comp.}
\textbf{1987}, \textit{49}, 91--103.
\url{https://doi.org/10.1090/S0025-5718-1987-0890255-3}.

\bibitem{ref-shuosher1988}
Shu, C.-W.; Osher, S. Efficient implementation of essentially
non-oscillatory shock-capturing schemes.
\textit{J. Comput. Phys.} \textbf{1988}, \textit{77(2)}, 439--471.
\url{https://doi.org/10.1016/0021-9991(88)90177-5}.

\bibitem{ref-einfeldt1988}
Einfeldt, B. On Godunov-Type Methods for Gas Dynamics.
\textit{SIAM J. Numer. Anal.} \textbf{1988}, \textit{25(2)}, 294--318.
\url{https://doi.org/10.1137/0725021}.

\bibitem{ref-vreugdenhil1994}
Vreugdenhil, C.B. \textit{Numerical Methods for Shallow-Water Flow};
Kluwer Academic Publishers: Dordrecht, The Netherlands, 1994.

\bibitem{ref-toro1994}
Toro, E.F.; Spruce, M.; Speares, W. Restoration of the contact surface
in the HLL-Riemann solver. \textit{Shock Waves} \textbf{1994},
\textit{4}, 25--34.
\url{https://doi.org/10.1007/BF01414629}.

\bibitem{ref-toro2001}
Toro, E.F. \textit{Shock-Capturing Methods for Free-Surface Shallow
Flows}; John Wiley \& Sons: Chichester, UK, 2001.

\bibitem{ref-gottlieb2001}
Gottlieb, S.; Shu, C.-W.; Tadmor, E. Strong Stability-Preserving
High-Order Time Discretization Methods.
\textit{SIAM Rev.} \textbf{2001}, \textit{43(1)}, 89--112.
\url{https://doi.org/10.1137/S003614450036757X}.

\bibitem{ref-leveque2002}
LeVeque, R.J. \textit{Finite Volume Methods for Hyperbolic Problems};
Cambridge University Press: Cambridge, UK, 2002.
\url{https://doi.org/10.1017/CBO9780511791253}.

\bibitem{ref-audusse2004}
Audusse, E.; Bouchut, F.; Bristeau, M.-O.; Klein, R.; Perthame, B.
A Fast and Stable Well-Balanced Scheme with Hydrostatic
Reconstruction for Shallow Water Flows.
\textit{SIAM J. Sci. Comput.} \textbf{2004}, \textit{25(6)},
2050--2065.
\url{https://doi.org/10.1137/S1064827503431090}.

\bibitem{ref-george2008}
George, D.L. Augmented Riemann Solvers for the Shallow Water
Equations Over Variable Topography with Steady States and
Inundation. \textit{J. Comput. Phys.} \textbf{2008},
\textit{227(6)}, 3089--3113.
\url{https://doi.org/10.1016/j.jcp.2007.10.027}.

\bibitem{ref-delestre2013}
Delestre, O.; Lucas, C.; Ksinant, P.-A.; Darboux, F.;
Laguerre, C.; Vo, T.N.T.; James, F.; Cordier, S. SWASHES: a
compilation of shallow water analytic solutions for hydraulic
and environmental studies. \textit{Int. J. Numer. Methods Fluids}
\textbf{2013}, \textit{72}, 269--300.
\url{https://doi.org/10.1002/fld.3741}.

\bibitem{ref-kh2d2025}
Herho, S.H.S.; Trilaksono, N.J.; Fajary, F.R.; Napitupulu, G.;
Anwar, I.P.; Khadami, F.; Irawan, D.E. kh2d-solver: A Python
Library for Idealized Two-Dimensional Incompressible
Kelvin--Helmholtz Instability.
\textit{Appl. Comput. Mech.} \textbf{2025}, \textit{19},
125--156. \url{https://doi.org/10.24132/acm.2025.1040}.

\bibitem{ref-waveatt2026}
Herho, S.H.S.; Anwar, I.P.; Khadami, F.; Ndruru, T.R.E.B.N.;
Suwarman, R.; Irawan, D.E. wave-attenuation-1d: An Idealized
One-Dimensional Framework for Wave Attenuation Through Coastal
Vegetation Using Numba-Accelerated Shallow Water Equations.
\textit{J. Theor. Appl. Mech.} \textbf{2026}, \textit{56},
89--102. \url{https://doi.org/10.55787/jtams.2026.1.AI00236}.

\bibitem{ref-sangkuriang2026}
Irawan, D.E.; Herho, S.H.S.; Pamumpuni, A.; Kartiko, R.D.;
Khadami, F.; Anwar, I.P.; Sujatmiko, K.A.; Handayani, A.P.;
Fajary, F.R.; Suwarman, R. An Open-Source Pseudo-Spectral
Solver for Idealized Korteweg--de Vries Soliton Simulations.
\textit{Water} \textbf{2026}, \textit{18}, 779.
\url{https://doi.org/10.3390/w18070779}.

\bibitem{ref-matplotlib}
Hunter, J.D. Matplotlib: A 2D Graphics Environment.
\textit{Comput. Sci. Eng.} \textbf{2007}, \textit{9(3)}, 90--95.
\url{https://doi.org/10.1109/MCSE.2007.55}.

\bibitem{ref-pandas}
McKinney, W. Data Structures for Statistical Computing in Python. In
\textit{Proceedings of the 9th Python in Science Conference (SciPy
2010)}; van~der~Walt, S., Millman, J., Eds.; Austin, TX, USA, 28
June--3 July 2010; pp.~56--61.
\url{https://doi.org/10.25080/Majora-92bf1922-00a}.

\bibitem{ref-numba}
Lam, S.K.; Pitrou, A.; Seibert, S. Numba: A LLVM-Based Python JIT
Compiler. In \textit{Proceedings of the Second Workshop on the LLVM
Compiler Infrastructure in HPC (LLVM-HPC 2015)}; Austin, TX, USA, 15
November 2015; pp.~1--6.
\url{https://doi.org/10.1145/2833157.2833162}.

\bibitem{ref-numpy}
Harris, C.R.; Millman, K.J.; van~der~Walt, S.J.; Gommers, R.;
Virtanen, P.; Cournapeau, D.; Wieser, E.; Taylor, J.; Berg, S.;
Smith, N.J.; Kern, R.; Picus, M.; Hoyer, S.;
van~Kerkwijk, M.H.; Brett, M.; Haldane, A.;
del~R\'{i}o, J.F.; Wiebe, M.; Peterson, P.;
G\'{e}rard-Marchant, P.; Sheppard, K.; Reddy, T.;
Weckesser, W.; Abbasi, H.; Gohlke, C.; Oliphant, T.E. Array
programming with NumPy. \textit{Nature} \textbf{2020},
\textit{585}, 357--362.
\url{https://doi.org/10.1038/s41586-020-2649-2}.

\bibitem{ref-scipy}
Virtanen, P.; Gommers, R.; Oliphant, T.E.; Haberland, M.;
Reddy, T.; Cournapeau, D.; Burovski, E.; Peterson, P.;
Weckesser, W.; Bright, J.; van~der~Walt, S.J.; Brett, M.;
Wilson, J.; Millman, K.J.; Mayorov, N.; Nelson, A.R.J.;
Jones, E.; Kern, R.; Larson, E.; Carey, C.J.; Polat, \.{I};
Feng, Y.; Moore, E.W.; VanderPlas, J.; Laxalde, D.;
Perktold, J.; Cimrman, R.; Henriksen, I.; Quintero, E.A.;
Harris, C.R.; Archibald, A.M.; Ribeiro, A.H.; Pedregosa, F.;
van~Mulbregt, P.; SciPy 1.0 Contributors. SciPy 1.0:
fundamental algorithms for scientific computing in Python.
\textit{Nat. Methods} \textbf{2020}, \textit{17}, 261--272.
\url{https://doi.org/10.1038/s41592-019-0686-2}.

\bibitem{ref-netcdf4python}
Whitaker, J.; Khrulev, C.; Huard, D.; Hamman, J.; May, R.;
Fernandes, F. \textit{netCDF4: Python Interface to the NetCDF C
Library}; Unidata Program Center, University Corporation for
Atmospheric Research: Boulder, CO, USA.
\url{https://unidata.github.io/netcdf4-python/}.

\bibitem{ref-pillow}
Clark, A.; Murray, A.; van~Kemenade, H. \textit{Pillow: The Python
Imaging Library Fork}.
\url{https://pillow.readthedocs.io/}.

\bibitem{ref-tqdm}
da~Costa-Luis, C.O.; Larroque, S.K. \textit{tqdm: A Fast,
Extensible Progress Bar for Python and CLI}.
\url{https://tqdm.github.io/}.

\end{thebibliography}
\end{document}